\documentclass[12pt,preprint]{emulateapj}
\usepackage{natbib}

\newcommand{\mic}{$\mu$m}
\newcommand{\spitzer}{{\it Spitzer}}

\slugcomment{Last edited: \today}

\shortauthors{P\'erez-Gonz\'alez et al.}

\begin{document}

\title{{\it SPITZER} VIEW ON THE EVOLUTION OF STAR-FORMING GALAXIES FROM z$=$0 TO z$\sim$3}

\author{Pablo G. P\'erez-Gonz\'alez\altaffilmark{1}, George H. Rieke\altaffilmark{1}, Eiichi Egami\altaffilmark{1}, Almudena Alonso-Herrero\altaffilmark{2,1}, Herv\'e Dole\altaffilmark{3,1}, Casey Papovich\altaffilmark{1}, Myra Blaylock\altaffilmark{1}, Jessica Jones\altaffilmark{1}, Marcia Rieke\altaffilmark{1}, Jane Rigby\altaffilmark{1}, Pauline Barmby\altaffilmark{4}, Giovanni G. Fazio\altaffilmark{4}, Jiasheng Huang\altaffilmark{4}, Christopher Martin\altaffilmark{5}}







\altaffiltext{1}{The University of Arizona, Steward Observatory, 933 N Cherry Av., Tucson, AZ85721}
\altaffiltext{2}{Instituto de Estructura de la Materia, CSIC, Serrano 119, E-28006 Madrid, Spain}
\altaffiltext{3}{Institut d'Astrophysique Spatiale, b\^atiment 121, Universit\'e Paris Sud, F-91405 Orsay Cedex, France}
\altaffiltext{4}{Harvard-Smithsonian Center for Astrophysics, 60 Garden Street, Cambridge, MA02138}
\altaffiltext{5}{California Institute of Technology, Caltech, Pasadena, CA91125}


\begin{abstract} 

We use a 24~\mic\, selected sample containing more than 8,000 sources
to study the evolution of star-forming galaxies in the redshift range
from $z=0$ to $z\sim3$. We obtain photometric redshifts for most of
the sources in our survey using a method based on empirically-built
templates spanning from ultraviolet to mid-infrared wavelengths. The
accuracy of these redshifts is better than 10\% for 80\% of the
sample. The derived redshift distribution of the sources detected by
our survey peaks at around $z=0.6-1.0$ (the location of the peak being
affected by cosmic variance), and decays monotonically from $z\sim1$
to $z\sim3$. We have fitted infrared luminosity functions in several
redshift bins in the range $0<z\lesssim3$. Our results constrain the
density and/or luminosity evolution of infrared-bright star-forming
galaxies. The typical infrared luminosity ($L^*$) decreases by an
order of magnitude from $z\sim2$ to the present. The cosmic star
formation rate (SFR) density goes as $(1+z)^{4.0\pm0.2}$ from $z=0$ to
$z=0.8$. From $z=0.8$ to $z\sim1.2$, the SFR density continues rising
with a smaller slope.  At $1.2<z\lesssim3$, the cosmic SFR density
remains roughly constant. The SFR density is dominated at low redshift
($z\lesssim0.5$) by galaxies which are not very luminous in the
infrared ($L_\mathrm{TIR}<10^{11}\,L_\sun$, where $L_\mathrm{TIR}$ is
the total infrared luminosity, integrated from 8 to 1000~\mic).  The
contribution from luminous and ultraluminous infrared galaxies
($L_\mathrm{TIR}>10^{11}\,L_\sun$) to the total SFR density increases
steadily from $z\sim0$ up to $z\sim2.5$, forming at least half of the
newly-born stars by $z\sim1.5$. Ultraluminous infrared galaxies
($L_\mathrm{TIR}>10^{12}\,L_\sun$) play a rapidly increasing role for
$z\gtrsim1.3$.

\end{abstract}


\section{INTRODUCTION}
\label{intro}

Infrared surveys are rapidly achieving comparable sample sizes and
areal coverage as deep ultraviolet (UV) and optical ones, providing an
important perspective on galaxy evolution. Ground-based measurements
plus the {\it Infrared Astronomical Satellite}~(IRAS) have
demonstrated that star-forming galaxies are strong infrared
sources. Although star formation in disks is also readily detected in
the UV or through optical emission-lines, nuclear starbursts are often
heavily obscured, making infrared measurements essential to probe them
(\citealt{1998ARA&A..36..189K}). Early ground-based photometry
\citep{1972ApJ...176L..95R} and IRAS \citep{1988ApJ...328L..35S} also 
revealed a population of massive galaxies in the local Universe with
extremely high rates of star formation
($\mathrm{SFR}>100\,\mathcal{M}_\odot\,\mathrm{yr}^{-1}$): the
ultraluminous infrared galaxies (ULIRGs). This violent star formation
is almost completely undetectable in the optical and UV part of the
spectrum due to huge attenuation by dust. Together with
lower-luminosity dust-embedded starbursts, this type of activity
accounts for up to 20\% of the local star formation.

The {\it Infrared Space Observatory}~(ISO) showed that dust-enshrouded
starbursts have undergone strong evolution from $z\sim1$ to $z=0$
\citep{2001A&A...378....1F}. From results in the sub-mm with SCUBA, it 
appears that the output of ULIRGs may dominate the energy density in
the Universe at $z\ge2$.  However, the limitations in resolution and
sensitivity of most of the ISO surveys have not allowed reliable
identifications of a sufficient number of infrared galaxies at $z<1$
to estimate robust luminosity functions. The limitations are even more
severe for probing the ULIRG population at higher redshifts.  At
$z\ge1$, the 8~\mic\, polycyclical aromatic hydrocarbons (PAH) band is
shifted out of the longest ISOCAM band at 15~\mic, and the longer
wavelength ISOPHOT bands have limitations in both sensitivity and
angular resolution. Thus, we have only a first-order vision of the
importance of infrared-bright galaxies at $0<z<1$ in the general
picture of galaxy evolution (e.g., what percentage of the total star
formation rate density is contributed by ULIRGs and what for
optical/UV selected star-forming galaxies?). We have even less
information at $z=1-3$, where we believe the co-moving SFR density
reaches a maximum
\citep{2001MNRAS.320..504S,2002ApJ...570..492L}, most of the stars in
galaxies were formed
\citep{2003ApJ...587...25D,2003ApJ...596..734C}, and dynamic
structures (bars, disks) start to have a role in galaxy evolution
\citep{1998MNRAS.295..319M}. 

\spitzer's band at 24~\mic\, encompasses PAH emission to $z>2$, making 
star-forming galaxies readily detectable
\citep{2004ApJS..154..130E,2004ApJS..154..170L}. This band has 
sensitivity an order of magnitude greater than the ISO $15$~\mic\,
band, and 16 times as many pixels. \spitzer\, also has bands at 70 and
160~\mic, with similar gains over ISO to those at 24~\mic. \spitzer\,
therefore provides for the first time the ability to survey large
fields on the sky to adequate depth to resolve the majority of the
far-infrared (FIR) background and to characterize the $z\ge1$
population of ULIRGs and starbursts. Moreover, MIPS\footnote{{\it
Multiband Imaging Photometer for Spitzer}} provides a link between the
population of objects being discovered in the sub-mm and mm, and the
UV/optical and near-infrared (NIR) wavelengths, giving us a key new
tool to understand galaxy evolution.

This work is part of a series of papers where we will demonstrate the
ability of \spitzer\, to unveil galaxy evolution in the $0<z\lesssim3$
redshift range through its IRAC\footnote{{\it Infrared Array Camera}
on {\it Spitzer}.} and MIPS instruments. In \citet{astro-ph/0502246}
and \citet{casey05} we investigate the processes governing the
evolution of star-forming galaxies from $z=0$ to $z\sim1$. In
\citet{emeric05}, we study the luminosity evolution of infrared-bright
sources up to $z\sim1$ using a sample of sources detected by
\spitzer\, in the mid-infrared (24~\mic), and the extensive dataset in
the Chandra Deep Field South. The paper uses a combination of
spectroscopic redshifts and photometric redshifts from the
COMBO17\footnote{Classifying Objects by Medium-Band Observations, a
spectrophotometric 17-filter survey} project
\citep{2004A&A...421..913W}. However, these redshift surveys identify
few galaxies at $z\gtrsim1$. In the present paper, we extend the
previously mentioned work to $1<z\lesssim3$. We develop a photometric
technique based on IRAC and deep optical photometry
\citep[see][]{2004ApJS..154...44H,2004ApJS..154..107W,
2004ApJS..154..170L}, that is able to obtain reliable redshifts for
virtually all the galaxies detected by MIPS up to $z\sim3$. With these
redshifts, we build luminosity functions in the mid-infrared (MIR) in
several redshift bins. This will allow us to study the evolution of
star-forming galaxies and constrain the SFR history of the Universe up
to $z\sim3$ using a homogeneously selected sample and a SFR estimator
not affected by dust attenuation.

This paper is organized as follows. Section~\ref{sample} presents the
observations carried out with \spitzer, and the ancillary data
compiled for this study. Section~\ref{tech} describes the technique
used to estimate the redshifts from broad-band photometry for our
sample of galaxies. Further details about this technique are given in
Appendix~\ref{ppp}. The main results on the photometric redshifts are
presented in Section~\ref{res1}. The luminosity function estimation
and fitting, constraints on the cosmic star formation rate density,
and a discussion of the contribution of galaxies with different SFRs
and masses to the total SFR density of the Universe will be presented
in Sections~\ref{lumf} through \ref{fin}. The method for estimating
the luminosity functions, which takes into account the photometric
redshift errors, is described in Appendix~\ref{montecarlo}. Finally,
the conclusions are outlined in Section~\ref{conclu}. Throughout this
paper, we use a cosmology with $\mathrm
H_{0}=70$~km\,s$^{-1}$\,Mpc$^{-1}$, $\Omega_{\mathrm M}=0.3$ and
$\Lambda=0.7$. All magnitudes refer to the AB system.

\section{THE SAMPLE}
\label{sample}

\subsection{\spitzer\, observations}
\label{spit}

The sample used in this paper is drawn from MIPS 24~\mic\,
observations of the Chandra Deep Field South (CDFS,
$\alpha=03^h32^m02^s$, $\delta=-27\arcdeg37\arcmin24\arcsec$, J2000)
and the Hubble Deep Field North (HDFN, $\alpha=12^h37^m57^s$,
$\delta=+62\arcdeg23\arcmin14\arcsec$, J2000). In each field, we used
the scan map mode to observe a rectangle of
$1.5\arcdeg\times0.5\arcdeg$ in each of the three MIPS wavelengths
(24, 70, and 160~\mic). The overlay zone covered with all three
channels is $1.0\arcdeg\times0.4\arcdeg$. To have the most ancillary
data for each 24~\mic\, selected source, we concentrated this work in
a smaller area around the COMBO17 \citep{2004A&A...421..913W},
GOODS\footnote{The Great Observatories Origins Deep Survey.} ACS
\citep{2004ApJ...600L..93G}, and ESO Imaging Survey (EIS,
\citealt{2002yCat..33790740A}) pointings in the CDFS case, and around
the GOODS ACS footprint in the HDFN. In both fields, we also obtained
IRAC data. The CDFS and HDFN were observed in the four IRAC channels
(at 3.6, 4.5, 5.8, 8.0~\mic) covering an area of
$1.0\arcdeg\times0.5\arcdeg$ in each field.


The reduction of the 24~\mic\, images was carried out with the MIPS
Data Analysis Tool \citep{2004PASP...DAT}. The final images had an
average exposure time of $\sim1400$~s. Source detection and photometry
were carried out with several tasks ({\it daofind}, {\it phot}, and
{\it allstar}) in the {\sc DAOPHOT} package in the Imaging Reduction
and Analysis Facility, IRAF\footnote{IRAF is distributed by the
National Optical Astronomy Observatories, which is operated by the
Association of Universities for Research in Astronomy, Inc. (AURA)
under cooperative agreement with the National Science
Foundation.}. Sources were detected in two passes to recover the
faintest sources, many of which are hidden by brighter
ones. Photometry was extracted for all the sources (from the two
passes) together to obtain the best possible results in crowded
regions. Given the large point spread function (PSF) of the MIPS
24~\mic\, channel (which produces very crowded images), all
measurements were made by PSF fitting. For sources of noticeable
extent, the measurement aperture was set accordingly. For the rest, a
circular aperture of size $\sim15\arcsec$ was utilized. We used an
aperture correction based on the theoretical PSF of MIPS to correct to
the total flux. The sky estimation was carried out in two steps, first
removing the large-scale variation (due to Zodiacal light) and then
measuring the background around each source.

IRAC images were reduced with the general \spitzer\, pipeline, and
then mosaicked. The average exposure time of these frames is
approximately 500~s. Source detection and photometry was carried out
with {\sc SEXTRACTOR} \citep{1996A&AS..117..393B}, using the same
procedure as \citet{2004ApJS..154...44H}. Special care was taken with
the deblending of sources, given the large density of objects and the
marked features of the IRAC PSF. Since the third and fourth IRAC
channels are less sensitive than the first and second ones, we tried
to obtain fluxes for the faintest sources in the former by carrying
out the detection in the latter and measuring photometry in all
bands. Photometry was performed using a small circular aperture
($3\arcsec$ in diameter) and an aperture correction was applied to get
the total flux (assumed to be the one corresponding to a circular
aperture of diameter 24.4$\arcsec$). The aperture corrections were
calculated from in-flight PSFs. For extended sources, a circular
aperture large enough to capture the total signal was used.

The catalogs for the \spitzer\, bands in both the CDFS and HDFN were
cut for sky regions where the IRAC/MIPS coverage was the deepest, and
other UV/optical/NIR data were available (see next Section). We
detected 4373 sources in the CDFS ($4\sigma$ above the sky level) in a
area of 665~arcmin$^2$; 4593 sources were detected in the HDFN in
517~arcmin$^2$ (above $3\sigma$ of the sky level\footnote{We used a
lower detection limit in the HDFN given the ultra deep optical and NIR
data that we had in this field. In comparison with the CDFS dataset,
limiting magnitudes for the HDFN images are $0.5-1.0$~mag
fainter. This allowed us to identify fainter MIPS sources which were
flagged as non-spurious because they were also detected in the optical
and/or NIR.}). In \citet{2004ApJS..154...70P}, we estimated that the
final catalogs are 80\% complete at $F_{24}=83$~$\mu$Jy.

\subsection{Ground based optical and near-infrared photometry}
\label{ground}

The \spitzer\, images were complemented with the extensive dataset
available for both the CDFS and HDFN. For the CDFS, we used the
publicly available optical images ($UU_pBVRI$) released by EIS
\citep{2002yCat..33790740A}, the optical fluxes from COMBO17 
\citep{2004A&A...421..913W},  the $RIz$ frames published by the
Las Campanas Infrared Survey \citep{1999ASPC..191..148M}, the HST/ACS
$bviz$ observations carried out by GOODS
\citep{2004ApJ...600L..93G}, the near-infrared $JK$ data released by
the EIS Deep Public Survey (EIS-DPS, \citealt{2001astro.ph..2300V}),
$JHK$ frames released by GOODS \citep{2004ApJ...600L..93G}, and the
$I$-band photometry and spectroscopic redshifts released by the
VIRMOS-VLT Deep Survey (VVDS, \citealt{2004A&A...428.1043L}). We also
used UV data taken with GALEX in two bands at 150~nm (FUV-band) and
230~nm (NUV-band). For the HDFN, the \spitzer\, data were complemented
with publicly available ultra-deep optical and NIR data spanning from
the $U$- to the $HK_s$-band ($UBVRIzHK_s$,
\citealt{2004AJ....127..180C}). We also used the $bviz$ images
published by GOODS for the central region in the HDFN. For all these
images, source detection and photometry were carried out with {\sc
SEXTRACTOR} \citep{1996A&AS..117..393B}. We refer the reader to
Appendix~\ref{ppp} for more detailed information about the data
compilation carried out for this paper.




\subsection{Redshift ancillary data}
\label{redancillary}

Spectroscopic redshifts for 1599 sources have been released by
\citet{2004A&A...428.1043L} for the CDFS (VVDS redshifts). In addition, 
COMBO17 observed this field with up to 17 medium- and broad-band
filters to obtain high quality photometric redshifts
\citep{2004A&A...421..913W}. For $R<24$, COMBO17 gives redshifts for
approximately 11,000 sources. A total of 425 galaxies in our sample in
the CDFS have VVDS spectroscopic redshifts (9\% of the sample), and
2118 (48\%) have COMBO17 photometric redshifts. In the HDFN, several
spectroscopic surveys have been or are being carried out. Most of the
spectroscopic redshifts have been compiled by the Team Keck Treasury
Redshift Survey (TKRS, \citealt{2004AJ....127.3121W}) and
\citet{2004AJ....127.3137C}. We have also used the photometric
redshifts found in \citet{1999ApJ...513...34F}. Out of the total
number of galaxies in the HDFN survey, 601 sources (13\%) have a
spectroscopic redshift.


\subsection{Merged catalogs}
\label{merging}

Merged catalogs in all the available bands were built by matching the
coordinates of the 24~\mic\, sources to one (the deepest) reference
optical band ($B$ band in the CDFS and $R$ band in the HDFN). A
$2\arcsec$ search radius was used. Within this search radius, multiple
identifications (i.e., multiple sources in the optical/NIR
corresponding to the same 24~\mic\, detection) were found for no more
than 7\% of the sources (in the deepest ground-based images). At this
low rate, we do not expect multiple identifications to bias our
results.

To measure the photometry, we determined the elliptical aperture (from
isophote fitting in the reference band) corresponding to 2.5 times the
Kron radius (which contains more than 95\% of the total flux of the
source, according to \citealt{1980ApJS...43..305K}). In all cases the
apertures were large enough to enclose the PSF profile. This aperture
was translated to all the other optical and NIR images. This procedure
allowed us to obtain integrated fluxes for each filter in matched
apertures, and to estimate the color properly for each source. For
sources not detected in the reference image, we used other optical/NIR
images as the reference (if possible). In the case of the IRAC and
MIPS bands, where the PSF is larger than the object images, the
integrated flux was assumed to be that obtained from PSF fitting.

\begin{figure}
\begin{center}
\includegraphics[angle=-90,width=9cm]{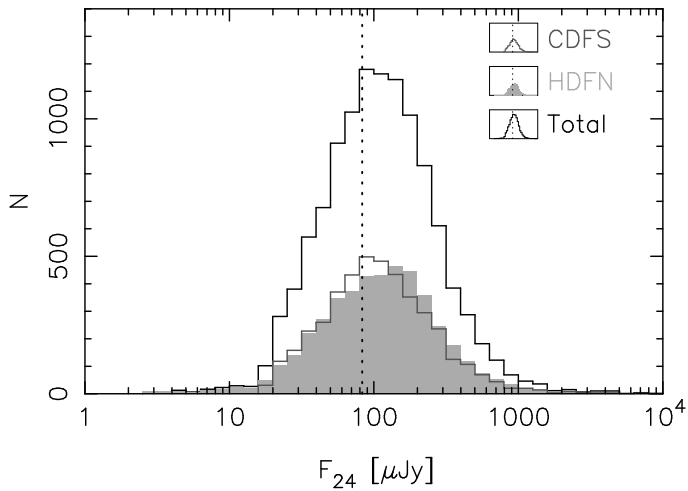}
\figcaption{\label{histf} MIPS 24~\mic\, flux distribution of the sources 
detected in the CDFS (open histogram limited by a gray line), HDFN
(filled gray histogram), and the total survey (open histogram limited
by a black line). The 80\% completeness flux level is shown at
$F_{24}=83$~$\mu$Jy \citep{2004ApJS..154...70P}.}
\end{center}
\end{figure}

Figure~\ref{histf} shows the 24~\mic\, flux distribution of sources in
our sample. We also depict the 80\% completeness level (83~$\mu$Jy).
For fluxes much lower than this value,
\citet{2004ApJS..154...70P} showed that an increasing fraction of the 
detected sources might be spurious. We tried to identify the real
24~\mic\, sources by cross-correlating the MIPS positions with
UV-to-MIR catalogs. A simulation of the source density in UV-to-MIR
images revealed that for a random position on the sky of the 24~\mic\,
image, there is a $\sim$20\% probability of having a counterpart in
one of the bluer bands within the search radius. However, for a
24~\mic\, source with identifications in at least three other bands,
the probability of a spurious association is almost negligible (less
than 3\%). Thus, only the 24~\mic\, sources detected in three more
additional bands were considered as real. Using this criterion, from
the 4373 sources selected at 24~\mic\, in the CDFS, 4257 (97\%) were
flagged as non-spurious detections. In the HDFN, we confirmed the
detection for 4385 sources (96\% of the total 4593 detections). 85\%
of the possibly spurious sources were below the 83~$\mu$Jy
completeness threshold.

\slugcomment{Please, plot this figure with the width of one column}
\placefigure{histf}

Of the 4257 sources in the final merged CDFS catalog, 96\% were
detected by IRAC in at least one channel. The 4\% remaining objects
were always near very bright sources which interfered with the
deblending algorithm. In the HDFN, the percentage is 89\% of MIPS
sources detected in at least one IRAC channel.

In the optical, an average of $\sim70$\% of the 24~\mic\, selected
sample is detected down to $B=24.7$, $V=23.8$, $R=23.7$. This
percentage rises to almost 90\% for $B=25.2$, $V=24.7$, and $R=24.4$
(always using the limiting magnitudes as defined in
Appendix~\ref{ppp}). These statistics mean that 30\% of our 24~\mic\,
selected sample is part of a population of IR-bright sources,
starburst or galaxies with active galactic nuclei (AGNs), that are
missed by UV/optical deep surveys such as COMBO17. Only with very deep
imaging ($R>25$) from large telescopes (8--10 meter class) can we
detect these infrared-bright sources. In the NIR, $\sim30$\% of the
sources are detected down to $J=22.5$, $K=21.8$ (EIS data). In the UV,
30\% of the sample is detected with the GALEX NUV channel, and 14\%
with the FUV filter.

As mentioned above, our sample selection requires at least four
identifications (at 24~\mic\, and three more bands). Most MIPS sources
are detected in at least one IRAC channel. However, to be included in
our sample they typically need to be also detected in at least one
optical band \footnote{They can also be detected only in three IRAC
bands, but IRAC channels three and four are less sensitive than the
other two, which causes very faint sources to be often detected only
in the two bluer IRAC bands.}. Therefore, extremely red optically
faint galaxies, which are expected to reside preferentially at high
redshift, may not be included in the sample. However, the number of
such optically faint ($R\gtrsim25$) 24~\mic\, sources is at most
3--4\% of the total sample (and some of these 24~\mic\, sources may be
spurious), not large enough to change our results significantly (e.g.,
the redshift distribution presented in Section~\ref{res1}).

\section{PHOTOMETRIC REDSHIFT TECHNIQUE}
\label{tech}

The redshift is one of the most important parameters to understand a
distant galaxy. Although it is best determined with spectroscopy,
because of the required cost in observing time, photometric methods
are increasingly used. Moreover, photometric redshifts are the only
approach for very faint samples of galaxies ($R>25$ or $I>24$), given
the sensitivity of the currently available spectrographs. There are
now a number of works on the technique of photometric redshifts
\citep[e.g.,][]{2000A&A...363..476B,2000ApJ...536..571B,2004PASP..116..345C,
2004A&A...421..913W}, and on the results obtained with them
\citep[see, e.g,][]{1996Natur.381..759L,1996ApJ...468L..77G,1997AJ....113....1S,
2003MNRAS.345..819R,astro-ph/0310038,2004MNRAS.353..654B}.

These works fully develop the use of optical and NIR data to obtain
photometric redshifts. In this wavelength range, there are a wide
variety of spectrophotometric models and templates
\citep[e.g.,][]{1980ApJS...43..393C,1999ApJS..123....3L,
2003MNRAS.344.1000B}, which have been tested by many authors and seem
to describe accurately the emission from all types of
galaxies. Optical methods suffer, however, from dust attenuation,
which can have a dramatic effect on the derived photometric redshifts.

\spitzer\, surveys have opened another window in the photometric redshift
possibilities. First, the galaxies detected by \spitzer\, might be
very different in their spectral energy distribution (SED) properties
from the typical UV/optical and even NIR based surveys. Second, IRAC
observations can go much deeper than ground-based NIR ones in a
shorter time. Therefore, one can use up to four more NIR and MIR bands
along with the optical ones. Although the optical bands often include
important redshift indicators, the additional bands increase the
confidence in the redshift determinations
\citep{1997ApJ...486L..11C,2003MNRAS.345..819R}. The gain in confidence 
results largely from the detection of the 1.6~$\mu$m spectral bump
(see \citealt{1988A&A...193..189J, 2002AJ....124.3050S,
2004ApJS..154..170L}). An issue is that models have not been developed
in the NIR and MIR as thoroughly as in the UV/optical. The reason is
twofold: 1) NIR spectroscopy is a relatively new capability (compared
to optical); and 2) at these wavelengths, although dust attenuation is
almost negligible, dust emission starts to dominate the integrated
spectrum of galaxies (in the continuum and in the PAH spectral
features). Consequently, one needs detailed radiative transfer
calculations to obtain models including both the stellar and dust
contributions. Although some progress has been made on this topic
\citep[e.g.,][]{1996MNRAS.282.1005C,1997ApJ...487..625G,
1999A&A...350..381D, 2003PASJ...55..385T}, the set of models available
in the literature is far from complete in the description of the
emission of all galaxies from the UV to MIR wavelengths.

The lack of a complete and reliable set of models describing the SED
properties of galaxies from the UV to the far-infrared and radio
wavelengths convinced us to use another approach to the problem. We
built empirical broad-band SEDs for galaxies of known redshift (the
spectroscopic sample introduced in Section~\ref{redancillary}). The
resulting templates were used to fit all the galaxies in the entire
sample. If we have a large enough number of previously known
redshifts, and if this training set of galaxies is representative of
the entire sample, we should be able to obtain reliable redshifts
\citep[see][]{1995AJ....110.2655C}. This method is similar to a neural
network technique in the sense that we use the same photometric data
(of galaxies with known redshift) to train the photometric redshift
algorithm \citep[see][]{2003MNRAS.339.1195F,2004PASP..116..345C}. A
detailed description of our photometric redshift technique is given in
Appendix~\ref{ppp}, jointly with a discussion on the reliability of
the redshift estimations.

The companion paper by \citet{emeric05} provides a different type of
test to our redshifts. It is based entirely on well-tested photometric
and spectroscopic redshifts up to $z\sim1.1$. We have compared the
results in that paper with those reported here. As will be pointed out
at appropriate places in the following sections, the agreement is
excellent and demonstrates that the outliers do not bias our
results. This agreement substantially increases our confidence in the
similar results we obtain at $z>1.1$, where there are few
spectroscopic or previous photometric redshifts for our galaxies (see
discussion in Appendices~\ref{ppp} and \ref{montecarlo}).

\subsection{Stars}

Given the high Galactic latitude of the fields and the MIR selection
of the sample, stars were not expected to be detected in our survey in
large numbers. In the final catalogs, only 5 sources in the CDFS and 8
in the HDFN were clearly identified with stars (less than 0.1\% of the
sample), based on the continuously decreasing SED in the NIR and
MIR. This result was also checked by studying the STAR\_CLASS
parameter from SEXTRACTOR for the galaxies detected in the
optical/NIR.

\subsection{AGNs}
\label{agns}

Galaxies with an AGN are expected to be bright in the MIR-FIR due to
the emission of the hot dust surrounding the central engine (heated by
x-ray and UV photons coming from the black hole). Indeed, IR surveys
with ISO and \spitzer\, are effective in detecting AGNs
\citep{2002A&A...383..838F,2002ApJ...568..470F,2004ApJS..154..155A,
2004ApJS..154..160R,2004ApJS..154..166L}. Based on the observed SEDs,
we have tried to identify the AGNs (at least the most extreme cases)
in our sample by selecting sources with monotonically rising spectra
from optical to MIR-FIR wavelengths following a power-law and lacking
any spectral feature (as traced by distinguishable changes in the
slope of the SED). The MIR-FIR emission of these sources should be
dominated by an AGN \citep{2004ApJS..154..155A,2004ApJS..154..160R},
and they probably contribute non-negligibly to the bright end of the
IR luminosity function. In addition, the photometric redshift
estimation for these objects is very uncertain due to the lack of
marked spectral features. Given that in this paper we are mainly
interested in the evolution of star-forming galaxies, we carried out a
first correction for the presence of AGNs in the survey by removing
from the sample the sources with monotonically rising spectra
\citep{almudena05}. Approximately 5\% of the galaxies in the total
sample are within this group. This percentage is lower than the
expected fraction of AGNs in MIR-FIR surveys, estimated in the range
$10-20$\% by several authors based on local samples
\citep{1993ApJS...89....1R}, x-ray identifications consistent with an
active galaxy \citep{2001AJ....122....1B,
2001ApJ...554..742H,2002A&A...383..838F, 2004ApJS..154..160R}, and
other methods \citep{2004AJ....127.3075L}. The sources we removed are
likely to be the most extreme cases (Type 1 obscured AGNs) within the
active galaxy population. The existence of star formation (co-existing
with the AGN) contributing significantly to the total IR luminosity of
these galaxies is not completely ruled out, but the high dust
temperatures necessary to obtain a power-law in the IRAC bands seem to
point to a predominant AGN. The luminosity functions and cosmic SFR
density calculations in the next Section were obtained from the
``AGN-purged'' sample. Further analysis of the data will be necessary
to estimate the importance of AGNs in \spitzer\, surveys with higher
reliability.

\section{RESULTS}

\subsection{Redshift distribution}
\label{res1}

\begin{figure*}
\begin{center}
\includegraphics[angle=-90,width=16cm]{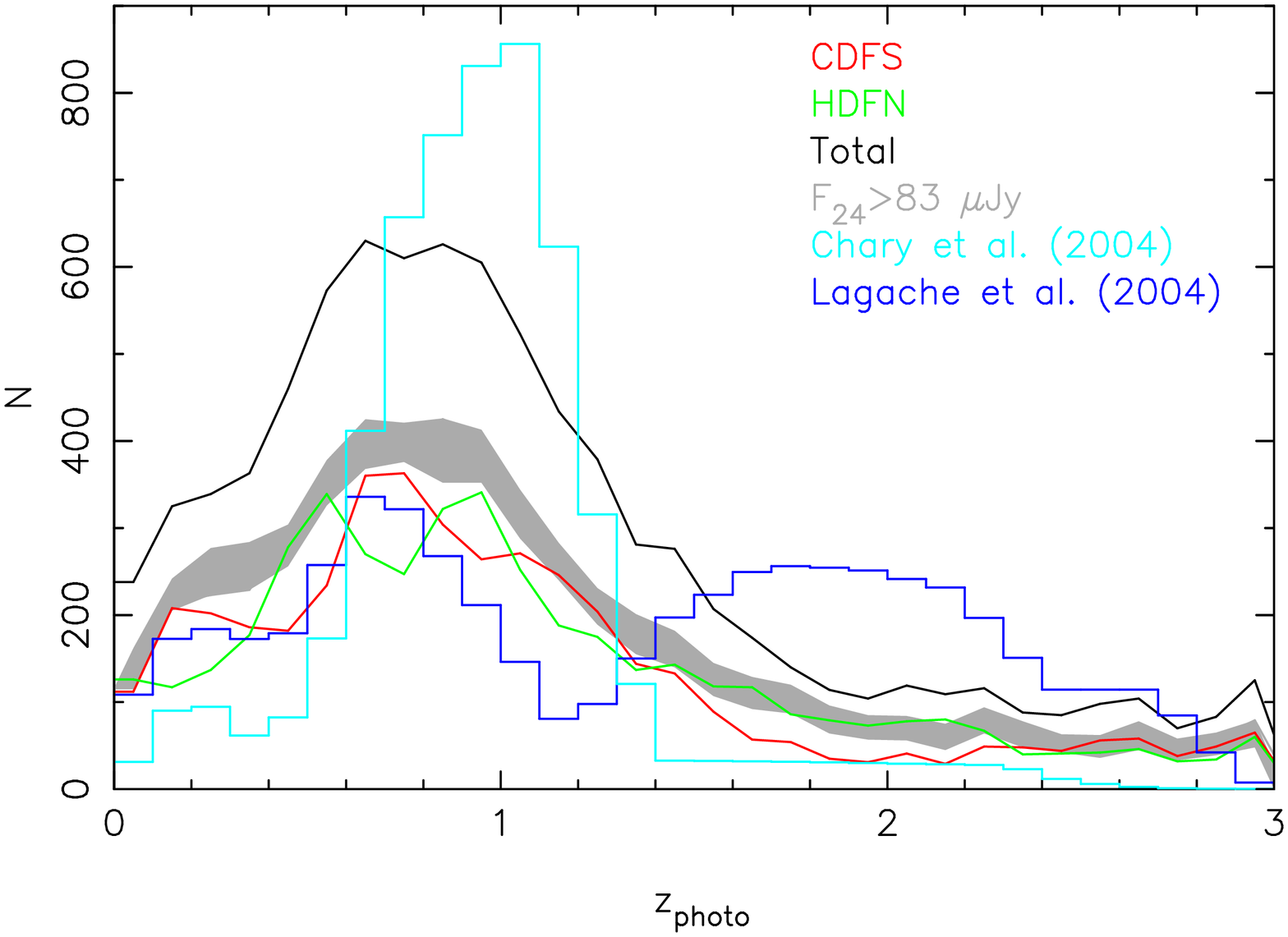}
\figcaption{\label{zdistri}Observed redshift distribution for the 24~\mic\, 
selected sources in the CDFS (red) and the HDFN (green). The combined
data for both fields and for all flux densities are plotted in
black. The predictions according to the models of
\citet{2004ApJS..154...80C} and \citet{2004ApJS..154..112L} for sources 
with 24~\mic\, flux densities larger than 83~$\mu$Jy are plotted in
cyan and blue, respectively. The shaded gray area shows the range of
results from our work (including the uncertainties in the photometric
redshifts) for the 83~$\mu$Jy limit (thus, this distribution is
directly comparable with the models).}
\end{center}
\end{figure*}

Figure~\ref{zdistri} shows the redshift distributions for the samples
of 24~\mic\, selected sources in the CDFS (in red), HDFN (in green),
and the total (black). These distributions result from the convolution
of the real redshift distribution of galaxies, the errors introduced
by the photometric redshift technique, and the (flux dependent)
detection curve of our survey. The cosmic variance between the two
fields is readily apparent. We find a redshift peak around $z\sim0.7$
in the CDFS, also seen in the COMBO17 optical survey
\citep{2004A&A...421..913W}. In the case of the HDFN, there seems to be
a density peak at $z\sim0.6$ and another one at $z\sim0.9$, which also
coincides with what was found by \citet{1999ApJ...513...34F} for the
WFPC2-HDF original field, and the spectroscopic results obtained by
the TKRS team. The apparent widths of these features ($\Delta
z\sim0.3$) support our estimate of the photometric redshift
errors. The curve for the total survey shows that the bulk of the
sources lie at $0.5<z<1.0$, just in the redshift range that ISO has
probed in recent years. However, the enhanced sensitivity of MIPS in
comparison with ISO has allowed us to detect a fainter population at
$z\lesssim1.4$ (see discussion below), and a significant number of
sources at $1<z<3$. In fact, almost half of the sample lies at $1<z<3$
(43\% of the sources in both fields). This is consistent with the
results obtained by
\citet{emeric05} using COMBO17 and VVDS redshifts for MIPS 24~\mic\, 
sources detected in the optical (55\% of the sources in the CDFS with
$R<24$ lying at $z<1$). Note also that 75\% of the sample is located
at $0<z<1.4$, for which Figure~\ref{comp_spec} and the discussion of
it directly confirm the reliability of our photometric redshifts.

\slugcomment{Please, plot this figure with the width of two columns}
\placefigure{zdistri}

Figure~\ref{zdistri} also shows model predictions (cyan and blue
histograms) for the 24~\mic\, MIPS detections above
$F_{24}=83$~$\mu$Jy
\citep{2004ApJS..154...80C,2004ApJS..154..112L}. These distributions are 
directly comparable with the gray curve, built from our sample with
the sources above the same flux cut (the shaded area depicts the
differences in the distribution given by the photometric redshift
errors). The general shape of the model distributions is roughly
similar to the observations for $z<1$. There are only small
differences in the position of the redshift peak, which is in any case
strongly affected by cosmic variance, as the curves for the CDFS and
the HDFN show. However, the density of sources in each redshift bin
below $z=1$ is very different from one model to the other, and from
the models to the data presented in this
paper. \citet{2004ApJS..154..112L} predict more sources at low
redshift ($z<0.4$) than
\citet{2004ApJS..154...80C}, and a less marked peak at $z\sim0.9$. Our
data lies between the models, presenting a $15-25$\% higher number
density than the prediction of
\citet{2004ApJS..154..112L} for $z<1$, and a 50\% lower density than
what the models by \citet{2004ApJS..154...80C} show at the peak.

The differences between the two models and the data are even larger at
$z>1$. According to \citet{2004ApJS..154...80C}, very few galaxies
should be detected at high redshift. On the contrary, the
\citet{2004ApJS..154..112L} model is bimodal, predicting that half of 
the infrared bright sources are at $z<1$ and half at $z>1$. The
percentage of sources lying at $z>1$ according to our photo-$z$ study
(43\%) seems to be consistent with this prediction, but most of them
are at $z\lesssim1.5$, whereas the model predicts that most are at
$z>1.4$. \citet{2004ApJS..154..112L} predict a steep decrease in the
number of sources detected at $z\sim1.2$ and a broad maximum around
redshift $z=1.8$ with a width of $\Delta z\sim1.0$. This maximum is
caused by prominent PAH features (at wavelengths from 6 to 10~\mic)
entering the MIPS 24~\mic\, filter as we move to higher redshifts. In
contrast, the observed redshift distribution shows a exponential decay
with very small and statistically irrelevant peaks at $z>1$. We do not
detect the predicted minimum in source counts at $z\sim1.2$, although
this feature could be washed out due to the errors inherent to the
photometric redshift technique. The general shape of the observed
distribution is closer to the prediction by
\citet{2004ApJS..154...80C}, although we obtain a substantially 
higher number density at $z\gtrsim1.4$.

The uncertainties in photometric redshifts are not able to explain the
difference between the \citet{2004ApJS..154..112L} model and our
results at high $z$. The modest portion of redshift outliers expected
from our estimates cannot be responsible for the inconsistency,
either. Compared with the models, we conclude there is a lower density
of sources at high redshift or that the sources that we are detecting
at $z>1$ do not present prominent PAH features in the
$6<\lambda<10$~\mic\, wavelength range (or both). These sources should
be very luminous (see Section~\ref{lumf}, Figure~\ref{zlum}, and
\citealt{2004ApJS..154..170L}), and PAH features could be absent or
hidden by a bright continuum or by silicate absorption, as some recent
luminosity dependent models predict \citep[see, e.g.,
][]{2001ApJ...556..562C,2002ApJ...576..159D}. Further analyses of the
SEDs of galaxies using the three MIPS wavelengths as well as the ISO
bands will be necessary to explore this result.

\begin{figure}
\begin{center}
\includegraphics[angle=-90,width=9cm]{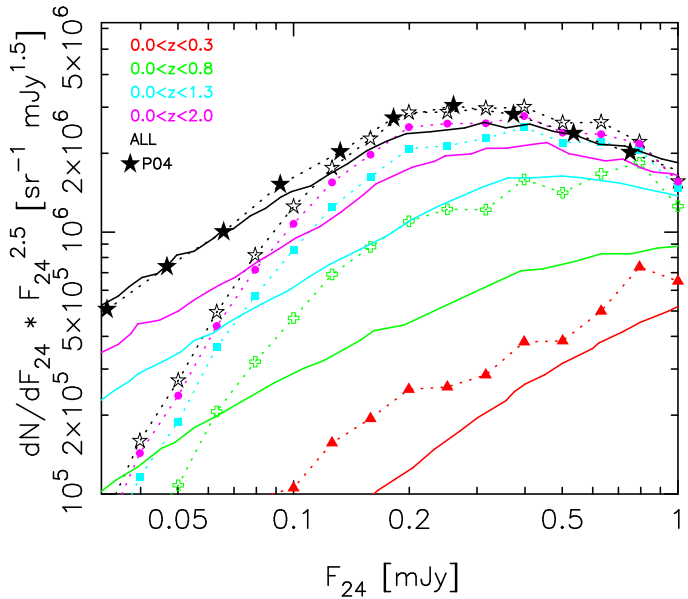}
\figcaption{\label{nc_z}Number counts at 24~\mic\, built for sources in 
several redshift ranges (not corrected for completeness). Symbols of
different colors (joined by dotted lines of the same color for
clarity) are used for each redshift range. The continuous lines show
the predictions from the \citet{2004ApJS..154..112L} models for each
redshift interval (in the same color as the data points).  Black
filled stars stand for the total number counts corrected for
completeness and presented in
\citet[][P04 in the figure]{2004ApJS..154...70P}.}
\end{center}
\end{figure}

\slugcomment{Please, plot this figure with the width of one column}
\placefigure{nc_z}

\citet{2004ApJS..154...70P}, \citet{2004ApJS..154...80C}, and
\citet{2004ApJS..154...66M} all presented number counts at 24~\mic. 
They all found that the peak in the differential number counts was
located at a fainter flux ($0.2-0.4$~mJy) than predicted by the models
based on ISO 15~\mic\, observations (roughly, at 1~mJy). In these
papers, it was argued that the difference implies the existence of a
previously undetected population of infrared-bright galaxies at
$z\sim1-3$. Using the photometric redshifts derived in this work, we
can study the contribution to the number counts of the galaxies in
different redshift bins, as shown in Figure~\ref{nc_z}. As we saw in
the previous Figure, the \citet{2004ApJS..154..112L} models
underpredict the number of sources at $z<1$, and overpredict the
number of galaxies above $z\sim1$. Our results seem to favor a
scenario where there is a strong evolution of infrared-bright sources
from $z=0$ to $z\sim1.0$, and then the evolution decelerates, stops or
even inverts \citep{2004ApJS..154...80C}. We will come back to this
issue when we present IR luminosity functions in
Section~\ref{lumfunc}.

\subsection{Infrared Luminosities and Star Formation Rates}
\label{lumf}

\begin{figure}
\begin{center}
\includegraphics[angle=-90,width=8cm]{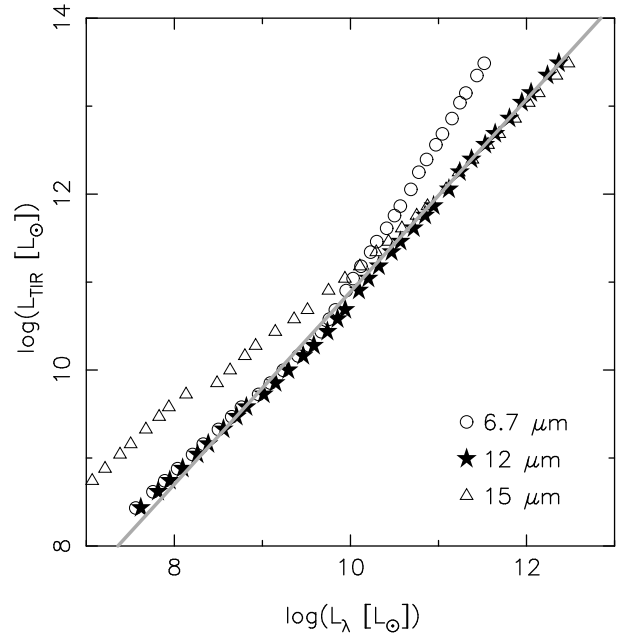}
\figcaption{\label{firlum}Correlation between 
the monochromatic luminosity at 6.7, 12, and 15~\mic\, and the TIR
emission (from 8 to 1000~\mic), according to the models by
\citet{2001ApJ...556..562C}. The best linear fit to the 12~\mic-to-TIR 
emission is also plotted
(Equation~\ref{12tofir}).}
\end{center}
\end{figure}

Models based on IRAS and ISO data on nearby galaxies can be used to
estimate the total infrared (TIR, integrated from 8 to 1000~\mic)
luminosity \citep[see, for
example,][]{1996ARA&A..34..749S,2001ApJ...556..562C,
2002ApJ...576..159D}. The most frequently used monochromatic fluxes to
estimate TIR luminosities are at 6.7, 12, 15~\mic\, (where the highest
quality ISO observations were carried out), and also 12, 25, 60, and
100~\mic\, (the IRAS bands). Measurements at 12~\mic\, have been shown
to be a useful estimator of the TIR emission for the luminosity range
that we are dealing with ($10^5\lesssim L_\mathrm{TIR}/L_\sun\lesssim
10^{13}$, see
\citealt{1989ApJ...342...83S,1995ApJ...453..616S,2001ApJ...556..562C}). This
conclusion is illustrated in Figure~\ref{firlum}, where we have
plotted the relationship between the monochromatic luminosity at 6.7,
12, and 15~\mic\, and the TIR emission according to the models of
\citet{2001ApJ...556..562C}. This Figure shows that the
6.7~\mic-to-TIR and the 15~\mic-to-TIR correlations present different
behaviors for different luminosity ranges. The 12~\mic\, data shows
the smallest scatter from a linear correlation. Similar results are
obtained with other models found in the literature, such as
\citet{1999A&A...350..381D} or \citet{2002ApJ...576..159D}.

We estimated TIR luminosities on the basis of the 12~\mic\, fluxes for
all the sources in the survey. This approach also allowed us to
compare with a vast number of papers in the literature based on IRAS
12~\mic\, observations of local galaxies. We estimated the rest-frame
monochromatic fluxes at 12~\mic\, by comparing the observed SEDs with
models of MIR-FIR emission. There are several sets of these models
available in the literature
\citep[e.g.,][]{2001ApJ...556..562C,2002ApJ...576..159D}. All of them 
combine an IR continuum with PAH emissions. The prominence of these
emissions depends on the TIR luminosity of the source. The set of
templates span a wide range of TIR luminosities (and, thus, PAH
emission properties). However, the models are only distinct for
wavelengths redder than $\sim3$~\mic\, in the case of
\citet{2001ApJ...556..562C}, and 10~\mic\, in the case of
\citet{2002ApJ...576..159D}, although the observed SEDs in real galaxies 
actually present very different shapes at bluer wavelengths (see
Figures~\ref{templas} and \ref{fits}). We chose the
\citet{2001ApJ...556..562C} template set (covering a wider range) and 
then compared these models with the observed SEDs for rest-frame
wavelengths redder than 3~\mic. The 12~\mic\, luminosity was assumed
to be the one corresponding to the model which best fitted the
observed data. For the sources at highest redshift, only one point
(the one for the 24~\mic\, observation) was available. In this case,
we estimated the 12~\mic\, luminosity by selecting the model best
fitting the luminosity measured by the 24~\mic\, channel.

\slugcomment{Please, plot this figure with the width of one column}
\placefigure{firlum}

The equations used to get the TIR emission from the 12~\mic\,
luminosity \citep{2001ApJ...556..562C}, and to obtain SFRs from the
TIR luminosity \citep{1998ARA&A..36..189K} are:


\begin{equation}
\label{12tofir}
log(L_\mathrm{TIR})=\log(0.89^{+0.38}_{-0.27})+1.094\times \log(L_\mathrm{12})
\end{equation}

\begin{equation}
\label{fir2sfr}
\mathrm{SFR}=1.71\times 10^{-10}L_\mathrm{TIR}
\end{equation}

\noindent where all the luminosities are in solar units, and the SFRs
in $\mathcal{M}_\sun\,\mathrm{yr}^{-1}$.

\begin{figure*}
\begin{center}
\includegraphics[angle=-90,width=16cm]{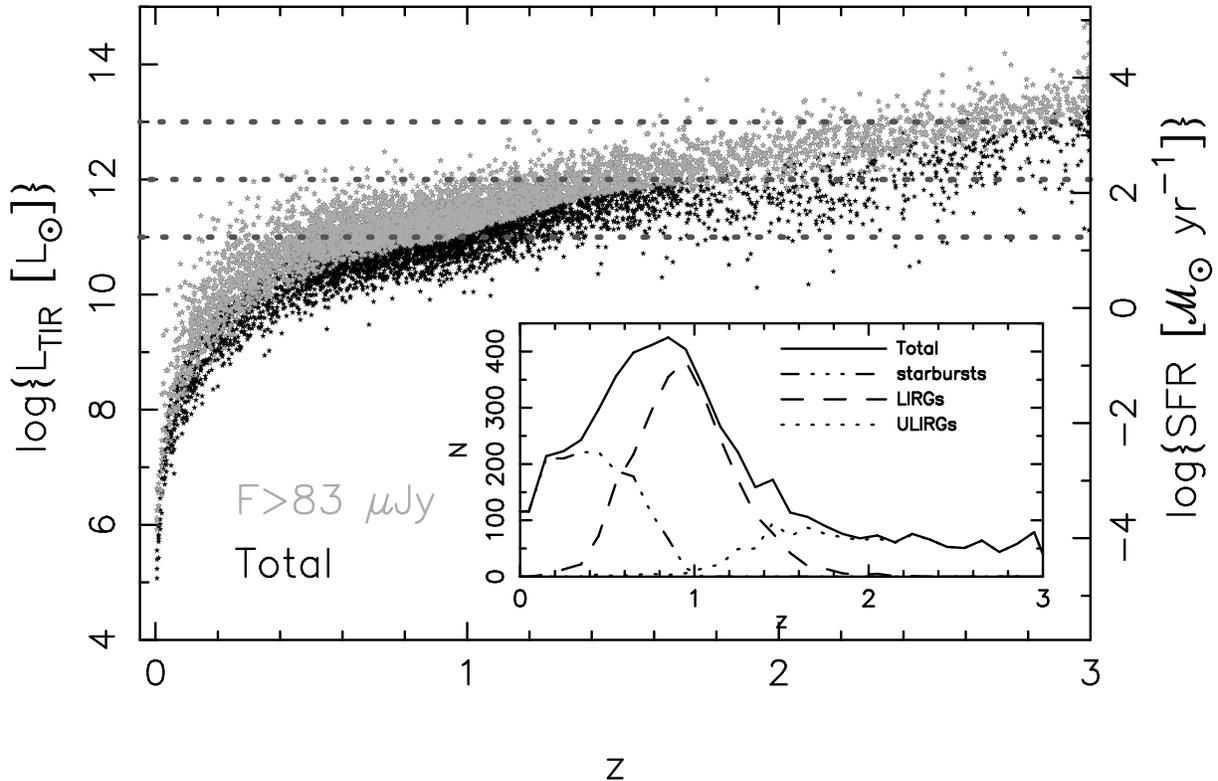}
\figcaption{\label{zlum}Selection effect on TIR luminosities of our 24~\mic\, 
survey for the entire sample (black stars) and sources with flux above
83~$\mu$Jy (gray stars).  The divisions for LIRGs
($10^{11}<L_\mathrm{TIR}<10^{12}\,L_\sun$), ULIRGs
($10^{12}<L_\mathrm{TIR}<10^{13}\,L_\sun$), and HyLIRGs
($L_\mathrm{TIR}>10^{13}\,L_\sun$) are marked with horizontal
lines. The inset plot shows the redshift distribution of sources with
flux above 83~$\mu$Jy divided into three luminosity ranges: starbursts
($L_\mathrm{TIR}<10^{11}\,L_\sun$), LIRGs
($10^{11}<L_\mathrm{TIR}<10^{12}\,L_\sun$), and ULIRGs
($L_\mathrm{TIR}>10^{12}\,L_\sun$).}
\end{center}
\end{figure*}

The estimation of the TIR emission from the monochromatic 12~\mic\,
luminosity, and the estimation of the 12~\mic\, luminosity itself, are
subject to uncertainties due to photometric redshift errors and the
dispersion of the different models (within the same library and from
one library to another; see
\citealt{2002ApJ...579L...1P,emeric05}). These uncertainties are
related to the heterogeneous dust properties observed in galaxies (PAH
strength and dust temperature), even for sources with the same
bolometric luminosity \citep{2004ApJS..154..178A}. One more caveat in
the calculation of the TIR luminosity is the questionable universality
of the relationship given in Equation~\ref{12tofir}, i.e., whether the
models derived from data for local galaxies apply to sources at higher
redshifts. In this sense, \citet{astro-ph/0502569} have analyzed a
sample of sources observed with both ISOCAM at 15~\mic\, and
\spitzer\, at 24~\mic\, and lying at a median redshift of $z\sim0.7$,
obtaining MIR colors that require the presence of PAHs for more than
half of their sample. Prominent PAHs have also been detected with
IRS\footnote{{\it Infrared Spectrograph for Spitzer.}} in galaxies up
to $z\sim3$ \citep{astro-ph/0502216}. Simulations of the observed IRAC
color-color diagrams \citep{2005ApJ...621..256S} and models of galaxy
evolution
\citep{2003ApJ...587...90X,2004ApJS..154...80C,2004ApJS..154..112L}
also seem to suggest the presence of PAHs in galaxies at all
redshifts. However, as we move to higher redshifts our survey is only
able to detect galaxies with high TIR luminosities (LIRGs and ULIRGs,
see Figure~\ref{zlum}). The MIR SEDs of these sources are probably
dominated by the continuum emission \citep{2001ApJ...556..562C}, and
consequently the PAH importance should decrease.

Based on all these arguments, we estimate that the 12~\mic\, and the
TIR luminosities calculated with Equation~\ref{12tofir} are accurate
within a factor of 2--3 for individual galaxies. In the following
discussion, we will avoid conclusions that would be affected by such
errors in the TIR luminosities. In future works, it is very desirable
to determine better MIR-FIR SEDs of galaxies at different redshifts to
improve the monochromatic to TIR relation. In fact, the uncertainty in
this correlation will dominate the errors quoted for the luminosities
of individual galaxies in the following sections. However, most of the
results to come (e.g., the luminosity functions and SFR densities)
depend on luminosities averaged over many galaxies. In this case, the
net errors should be reduced substantially.



Figure~\ref{zlum} shows the TIR luminosities of sources in our survey
as a function of redshift (black points). Gray points represent
sources with fluxes above the 80\% completeness limit (83~$\mu$Jy). A
sharp detection limit is seen. Very few galaxies are below this sharp
limit, probably most of them being photometric redshift outliers. This
plot also shows that we are able to detect galaxies with moderate star
formation (starburst galaxies with SFRs of a few
$\mathcal{M}_\sun\,\mathrm{yr}^{-1}$) up to $z\sim1$. Very few ULIRGs
are detected in this redshift range due to a low number density and/or
an insufficient area coverage. Above $z=1$ and up to $z\sim2$,
approximately half of the sources we are detecting present SFRs
typical of infrared-luminous galaxies (LIRGs with a few tens
$\mathcal{M}_\sun\,\mathrm{yr}^{-1}$), and half of them are ULIRGs
(SFR$\gtrsim$100~$\mathcal{M}_\sun\,\mathrm{yr}^{-1}$). Above
$z\sim2$, the Hyper-luminous infrared galaxy population (HyLIRGs,
$L_\mathrm{TIR}>10^{13}\,L_\sun$ or
SFR$\gtrsim$1700~$\mathcal{M}_\sun\,\mathrm{yr}^{-1}$) starts to be
detected. As expected, at $z\sim3$ the detection limit reaches only
HyLIRGs.

The inset of Figure~\ref{zlum} presents the contribution of galaxies
with different TIR luminosities to the redshift distribution presented
in Figure~\ref{zdistri}. This Figure is not corrected for
completeness, i.e., the distributions are affected by the selection
effects of our survey. A similar figure accounting for completeness
effects is shown in Figure~\ref{evollum}. The inset of
Figure~\ref{zlum} shows that starburst galaxies (defined as the ones
with $L_\mathrm{TIR}<10^{11}\,L_\sun$) dominate our survey for
redshifts below $z\sim0.5$, and are not detected beyond $z\sim1$. The
distribution for LIRGs shows a clear evolution from $z=0$, where very
few are detected, to $z\sim0.4$, where the number of detected LIRGs
starts to rise rapidly. At $z\sim1$, all the sources in the survey are
LIRGs. The curve for ULIRGs clarifies the statement in the previous
paragraph about the dominance in our survey of this kind of sources at
$z>2$.

\slugcomment{Please, plot this figure with the width of two columns}
\placefigure{zlum}

\subsection{Mid-infrared Luminosity functions}
\label{lumfunc}

We constructed and fitted luminosity functions at 12~\mic\, to study
the evolution of the total infrared output of the population of
galaxies as a function of redshift, and to put constraints on the
evolution of parameters such as the typical TIR luminosity of galaxies
($L^*$). The luminosity functions are weakly constrained at the
highest redshifts. Toward low luminosities, the achievable sensitivity
becomes an increasingly severe limitation, while our surveyed area is
inadequate to include rare, very high-luminosity objects. To estimate
the resulting systematic errors, we have used a variety of approaches
to fit luminosity functions to the data. First, we fitted a standard
form of luminosity function, allowing parametric adjustment of the
density normalization, the slope at faint luminosities, and $L^*$ to
minimize a $\chi^2$ likelihood estimator. Second, we used a variety of
functional forms determined for nearby galaxies and forced fits to the
data at different redshifts. We were careful in selecting the various
fitting approaches to include cases that would provide upper and lower
limits to the TIR luminosity density of the galaxy population,
allowing us to test our conclusions.

For all the luminosity function construction, we divided the sample
into redshift bins selected to provide adequate numbers of galaxies to
constrain the fits well. For $0<z<1$, we used five equal intervals
($\Delta z=0.2$). We used four additional intervals for $1<z<2.6$
($\Delta z=0.4$). No estimations were made for $z>2.6$, given the
small range of luminosities in our sample at such high redshift. The
estimation of the luminosity function was carried out using a stepwise
maximum-likelihood technique \citep[SWML,][see also
\citealt{1997AJ....114..898W}]{1988MNRAS.232..431E}, modified to take
into account the uncertainties in the photometric redshifts. The
procedure is described in Appendix~\ref{montecarlo}
\citep[see also][]{2003ApJ...586..745C}. The results were also 
checked using the $V/V_\mathrm{max}$ method
\citep{1968ApJ...151..393S,1973ApJ...186..433H}. 

\begin{figure}
\begin{center}
\includegraphics[angle=-90,width=8cm]{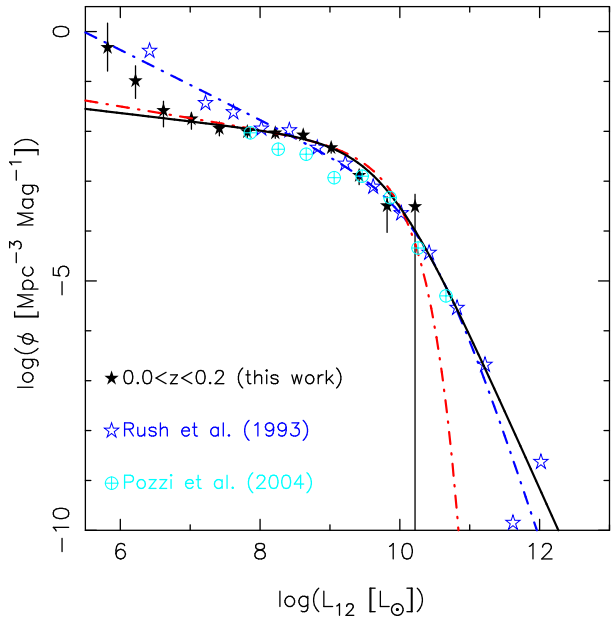}
\figcaption{\label{locallf} Local luminosity function at 12~\mic\, 
estimated by several authors using IRAS, ISO, and {\it Spitzer}
data. Black filled stars show the results from this work (obtained
with the entire sample of galaxies in the CDFS and the HDFN). Error
bars are the associated 1-$\sigma$ uncertainties based on Poisson
statistics \citep{1976ApJ...203..297S} and the propagation of
photometric redshift errors (see Appendix~\ref{montecarlo}). Our
results are compared with those achieved with IRAS data (for
non-active galaxies) by \citet[][ blue open
stars]{1993ApJS...89....1R} and those obtained with ISO observations
by \citet[][cyan crossed circles]{2004ApJ...609..122P}. The best fit
of our results to a \citet{1976ApJ...203..297S} function is plotted in
red, the fit to two power-laws (including the points from
\citealt{1993ApJS...89....1R} at $L_{12}>10^{10}\,L_\odot$) is plotted
in black (in both cases, taking into account the luminosity function
errors), and the double power-law fit given by
\citet{1993ApJS...89....1R} is plotted in blue.}
\end{center}
\end{figure}

\begin{figure*}
\begin{center}
\includegraphics[angle=-90,width=12cm]{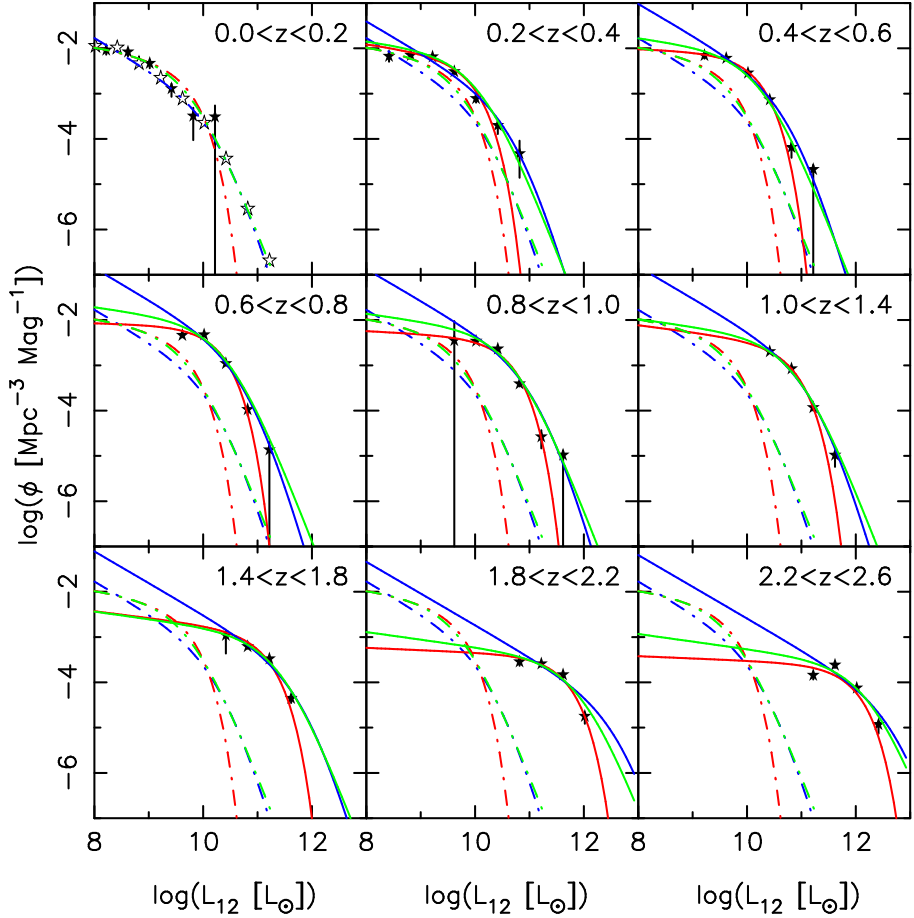}
\figcaption{\label{lf0t}Luminosity functions at 12~\mic\, for galaxies 
in several redshift ranges for both the CDFS and HDFN (black filled
stars in all the panels). For the sake of completeness and clarity, we
reproduce the 12~\mic\, local luminosity function given in
Figure~\ref{locallf} using smaller scales. The best fit to a
\citet{1976ApJ...203..297S} function is plotted in red, the fit to
two power-laws given by \citet{1993ApJS...89....1R} is plotted in
blue, and our own double power-law fit is plotted in green (all with
dot-dashed lines). The three fits of the local luminosity function are
also shown in all the other panels (with the same color and line
style). The continuous lines in the other panels show the best fits
for each of the approaches described in the text: independent fit to a
Schechter function (SCHLF), evolution of the local luminosity function
given by \citet[][RUSHLF]{1993ApJS...89....1R}, and evolution of the
local luminosity function obtained in this work (OWNLF). Red
continuous lines show the best SCHLF fit. Blue continuous lines show
the best RUSHLF fit for each redshift interval after applying a
density plus luminosity evolution. Green continuous lines show the
best OWNLF fit for each redshift interval after applying a density
plus luminosity evolution. Note that a pure luminosity evolution would
give practically the same results (i.e., we are not able to break the
degeneracy between the pure luminosity and the density plus luminosity
evolution).}
\end{center}
\end{figure*}

Once we had estimated the luminosity functions with the modified SWML
method, we carried out a variety of fits (see Figures~\ref{locallf}
and \ref{lf0t}). For the parametric fitting (SCHLF fit from now on),
we used a \citet{1976ApJ...203..297S} function\footnote{The functional
form is:
$\phi(L)dL=\frac{\phi^*}{L^*}\left(\frac{L}{L^*}\right)^\alpha
e^{-\frac{L}{L^*}}dL$} because it includes only three free parameters
(normalization $\phi^*$, faint-luminosity slope $\alpha$, and typical
luminosity $L^*$), in comparison with the four parameters in other
parametrizations (such as a double power-law). Each fit at each
redshift interval was independent from the others.

The Schechter parametrization is commonly used for UV/optical
luminosity functions. In our case, it fits the data points well for
$z>0.2$ with a minimum of free parameters. It is likely that the true
luminosity function is more similar to a double power-law, as many
authors have shown for the local infrared galaxy population
\citep[][among others]{1986MNRAS.219..687L,
1990MNRAS.242..318S,1993ApJS...89....1R,2001MNRAS.322..262S,
2003ApJ...587L..89T}. Our results in the local Universe also seem to
fit better to a double power-law, once you complement them with the
data from other works at the bright-end (see Figures~\ref{locallf} and
\ref{lf0t} and the discussion below). However, within the limited
luminosity range of our data for $z>0.2$ (where we can only estimate
the number density of sources for $L_{TIR}\gtrsim10^9\,L_\sun$), there
is little difference between a Schechter curve and a double
power-law. The extra parameter required in the double power-law is
therefore not well justified for our fits.

We also carried out a second set of fits, using two forms of double
power-law functions \footnote{The functional form used is:
$\phi(L)dL=\phi^*\left(\frac{L}{L^*}
\right)^{1+\alpha}\left(1+\frac{L}{L^*\beta}\right)^\beta dL$ 
(\citealt{1986MNRAS.219..687L} and
\citealt{1993ApJS...89....1R}).}: 
1) we used the local luminosity function derived by
\citet{1993ApJS...89....1R}; this fit (RUSHLF fit from now on)
presents a much steeper faint-end slope than the local SCHLF, and a
less steep slope at high luminosities (see Figure~\ref{locallf} and
the discussion of it); and 2) we used our own derivation of the local
luminosity function (OWNLF from now on), that presents an almost flat
faint-end slope (very similar to the SCHLF one) and is practically
identical to the local RUSHLF at high luminosities (see the discussion
below). The comparison of the three different fits to the data lets us
test directly the contribution of high luminosity galaxies to the
overall luminosity density of the population. With the high luminosity
shape fixed at the double power-law fit, the results as a function of
$\alpha$ also let us test the contribution of low infrared luminosity
galaxies to the overall output of the population. The probed values of
$\alpha$, from $\alpha=-1.7$ corresponding to the local RUSHLF fit to
$\alpha\sim-1.0$ corresponding to the SCHLF or to our OWNLF fit, are
typical for local infrared-selected galaxies
\citep{1993ApJS...89....1R, 1998ApJ...500..693F,1998ApJ...508..576X,
2004ApJ...609..122P}. This range is also very similar to the values
found at other wavelengths for different SFR estimators and at
different redshifts (see
\citealt{2004ApJ...615..209H} and references therein, and also
\citealt{2002ApJ...567..672C}, \citealt{2003ApJ...593L...1P},
\citealt{2004ApJ...606L..25B}, and \citealt{2004A&A...421...41G,
2004ApJ...616L..83G}).

\begin{figure*}
\begin{center}
\includegraphics[angle=-90,width=5.9cm]{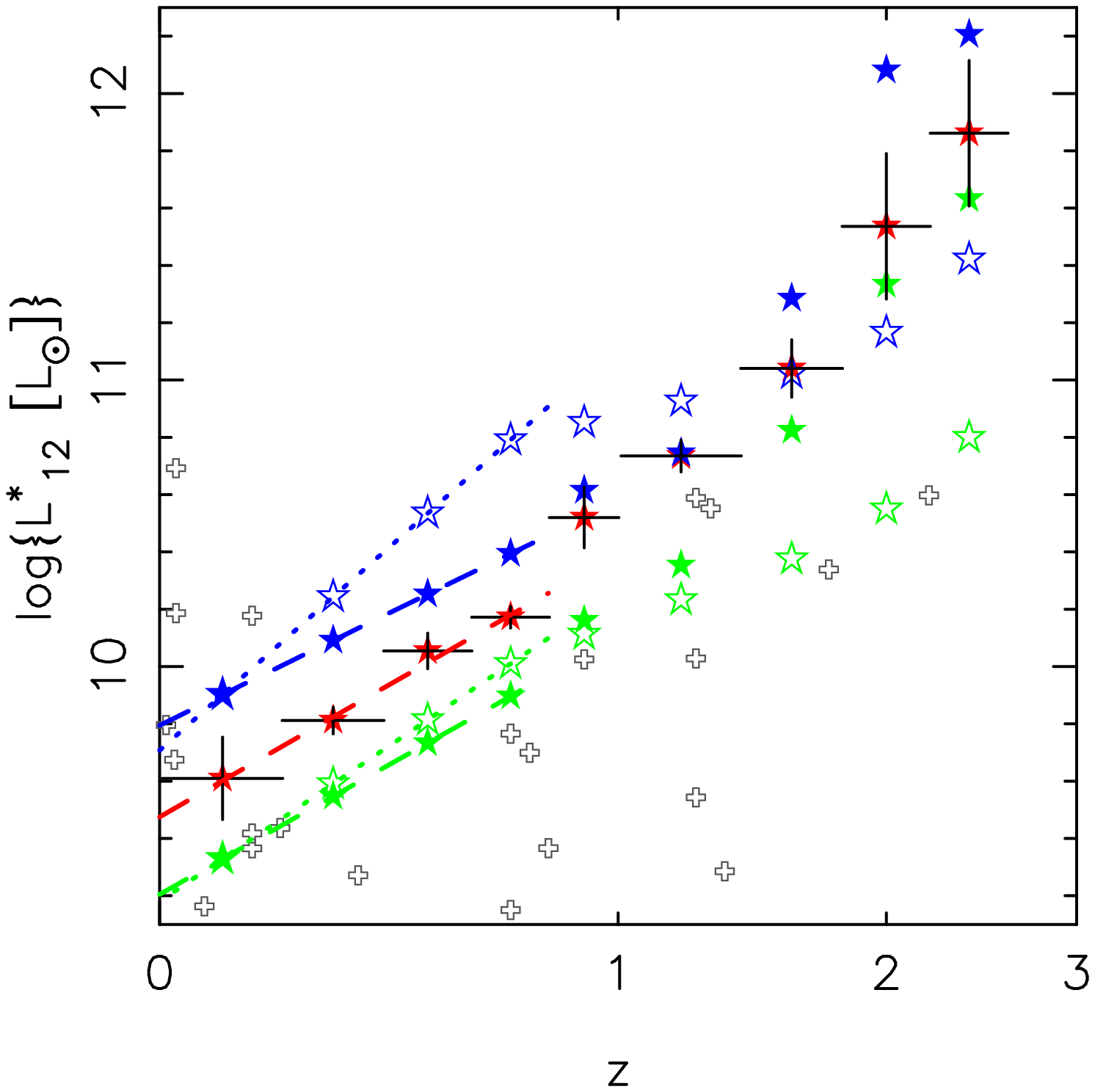}
\includegraphics[angle=-90,width=5.9cm]{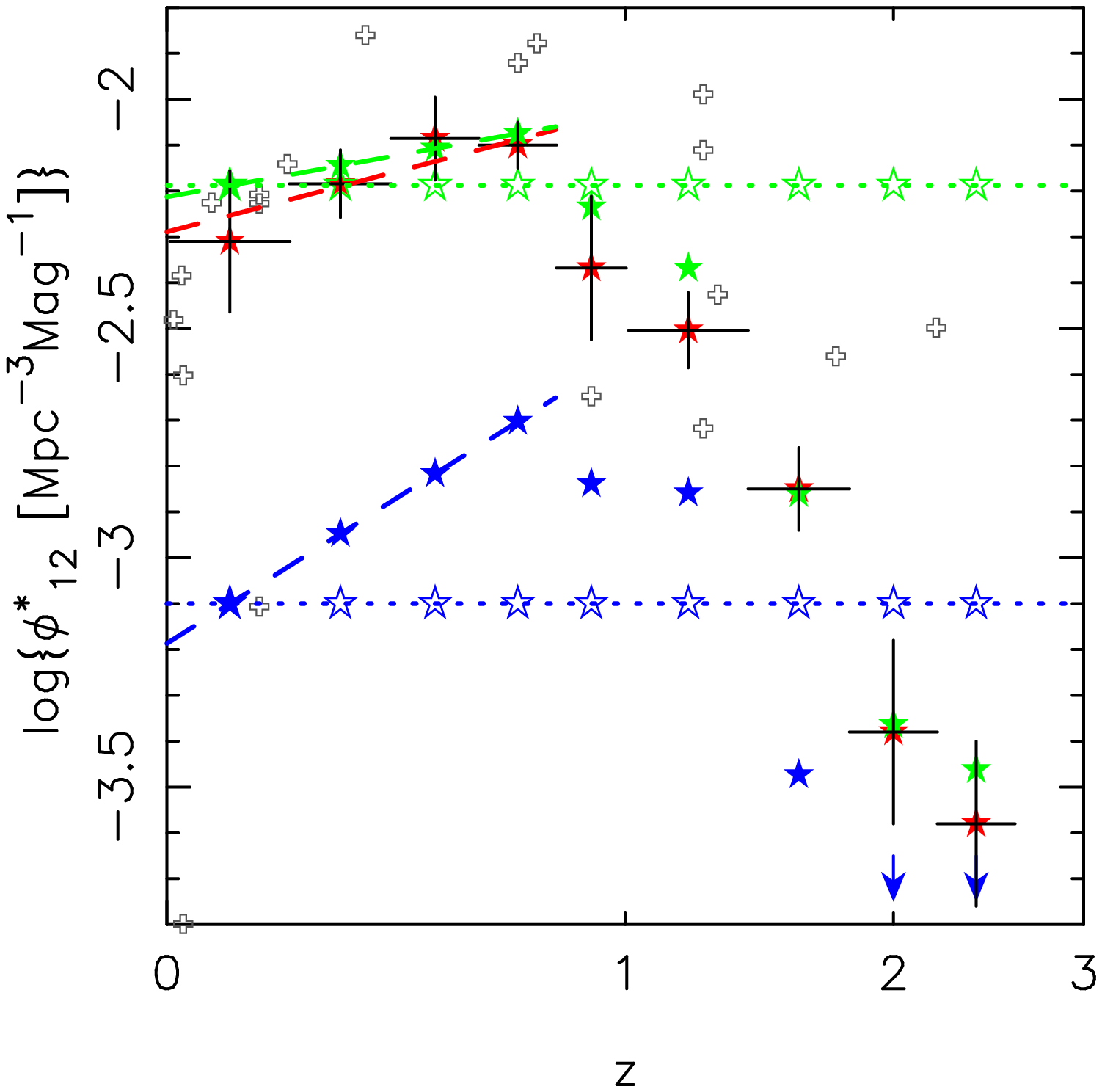}
\includegraphics[angle=-90,width=5.9cm]{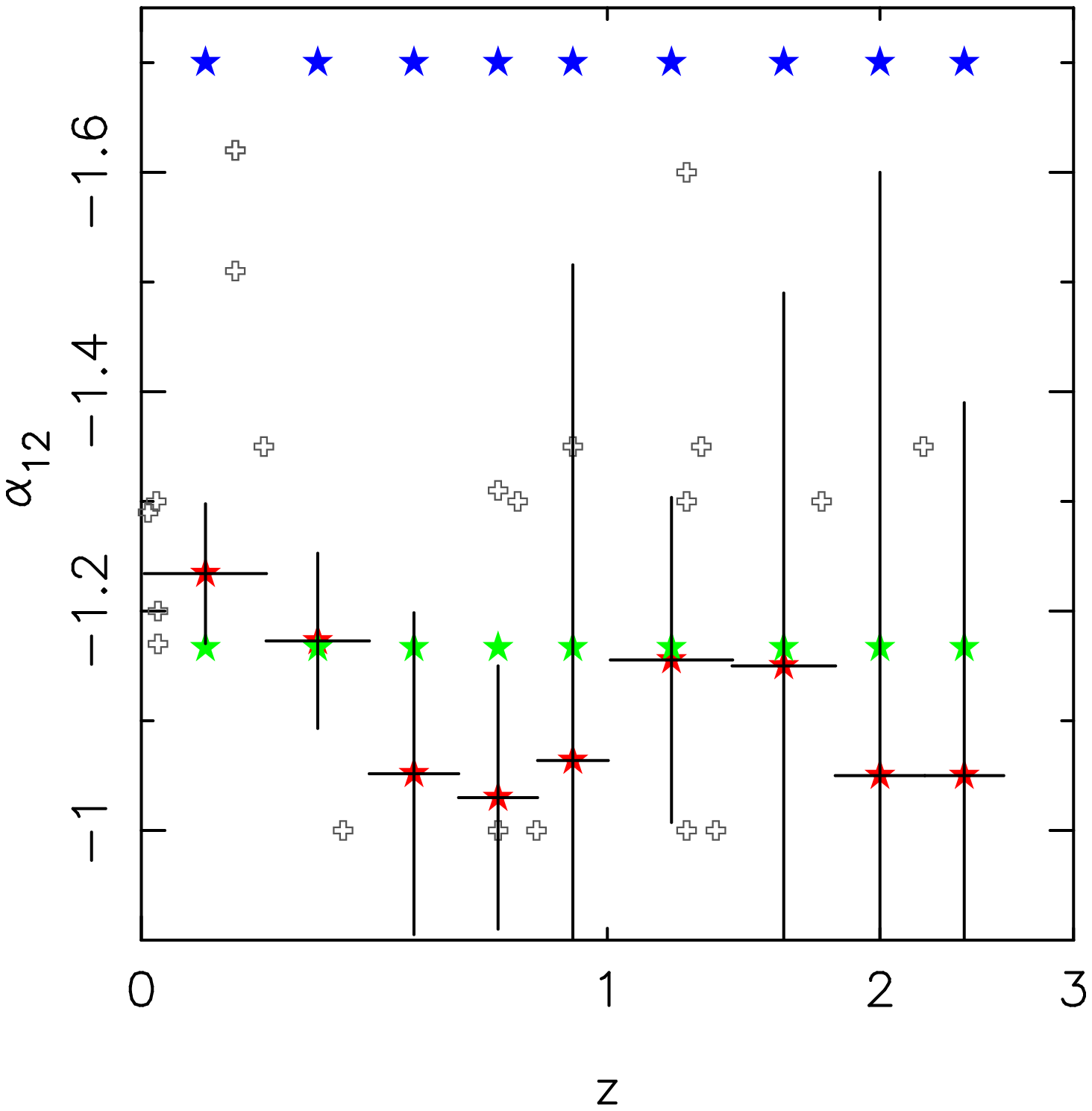}
\figcaption{\label{evol}Evolution of the 12~\mic\, luminosity function 
parameters from $z=0$ to $z\sim3$. For all panels, the results from
the SCHLF, RUSHLF, and OWNLF fits are plotted in red, blue, and green,
respectively. Open stars refer to the results for a pure luminosity
($L$) evolution, while filled stars refer to a combined luminosity
plus density evolution ($L+D$). The best fits for the evolution of
$\L^*$ and $\phi^*$ at $z<0.8$ are plotted with dotted and dashed
lines for $L$ and $L+D$ evolutions, respectively, with the same color
code as the points for the three sets of fits. Gray symbols show some
comparison values for the three parameters (for a Schechter
parametrization) extracted from the literature and based on fits of
luminosity functions built with samples of star-forming galaxies
selected with different SFR estimators (\citealt{1995ApJ...455L...1G,
1997ApJ...486L..11C,1998ApJ...495..691T,
1999AJ....118..603C,1999ApJ...519....1S,1999ApJ...519L..47Y,
2000A&A...360..463M,2000A&A...362....9M,2000AJ....120.2843H,
2000MNRAS.312..442S,2002AJ....124.1258W,2002ApJ...570L...1G,2002MNRAS.329..227S,
2002MNRAS.330..621S,2002MNRAS.337..369T,2003ApJ...589..704T,2003ApJ...591..827P};
all these estimations were compiled by
\citealt{2004ApJ...615..209H}, from which we extracted the SFR calibrations to 
convert all the $L^*$ values based on different estimators to
$L_{12}^*$). The evolution laws plotted in the Figure for a $L+D$
scenario are: $L^*$ evolves as $(1+z)^{3.1\pm0.5}$ and $\phi^*$
evolves as $(1+z)^{0.9\pm0.6}$ for the SCHLF fitting (red dashed
lines); $L^*\propto(1+z)^{2.6\pm1.1}$ and
$\phi^*\propto(1+z)^{2.1\pm0.6}$ for RUSHLF (blue dashed lines); and
$L^*\propto(1+z)^{3.0\pm0.3}$ and $\phi^*\propto(1+z)^{0.6\pm0.2}$ for
the OWNLF case (green dashed lines). For a pure $L$ evolution:
$L^*\propto(1+z)^{4.7\pm0.3}$ for RUSHLF (blue dotted line); and
$L^*\propto(1+z)^{3.6\pm0.3}$ for OWNLF (green dotted line).}
\end{center}
\end{figure*}

Most contemporary models of galaxy evolution assume that the observed
change of the infrared luminosity function with redshift can be
expressed through a number density and/or a luminosity evolution of
the local luminosity function. This means that the shape of the
luminosity function is conserved. The evolution is normally
parametrized with a $(1+z)^n$ law affecting the vertical axis (number
density), and another power-law making the luminosity function slide
along the horizontal axis (luminosity evolution). In the double
power-law set of fits, we allowed either a pure luminosity evolution
(of the local luminosity function) following a $(1+z)^{n_L}$ law ($L$
case), or a pure number density evolution following a $(1+z)^{n_D}$
law ($D$ case), or a combination ($L+D$ case), to test the
dependencies of the results on the assumed evolutionary behavior. The
results of the different fits are illustrated in Figures~\ref{locallf}
through \ref{evollum}. The recovered luminosity function parameters
jointly with the derived luminosity densities $\rho_{L_{12}}$ (for each
set of fits) are given in Tables~\ref{lfpars_sch} and
\ref{lfpars_pow}.

\slugcomment{Please, plot this figure with the width of one column}
\placefigure{locallf}

Figure~\ref{locallf} shows the local 12~\mic\, luminosity function
built with all the galaxies in our CDFS and HDFN surveys and the best
fits to it. Our results are compared with those obtained by
\citet{1993ApJS...89....1R} using  IRAS data and those achieved by 
\citet{2004ApJ...609..122P} using ISO
observations \citep[see also][]{1998ApJ...500..693F,
1998ApJ...508..576X}. Given that we only detect one galaxy with
$L_{12}>10^{10}\,L_\sun$, we assumed Rush et al. points in this regime
to obtain the OWNLF fit. For luminosities $10^8\lesssim
L_{12}\lesssim10^{10}\,L_\sun$, our results are very close to the IRAS
estimation, which presents slightly larger number densities than
ISO. For $L_{12}\lesssim10^8\,L_\sun$, our results are consistent with
a rather flat luminosity function down to
$L_{12}\sim10^{6.5}\,L_\sun$, and even below, in contrast to the IRAS
results. We find $\alpha=-1.23\pm0.06$ for the SCHLF fit
($\alpha=-1.17\pm0.07$ for the OWNLF fit), which is very close to the
results obtained for the local luminosity function of star-forming
galaxies using other SFR estimators (see, e.g.,
\citealt{2002MNRAS.330..621S}, \citealt{2003ApJ...591..827P}, 
\citealt{2005ApJ...619L..15W},  \citealt{2005ApJ...619L..31B}, 
and also Figure~\ref{evol}).


Figure~\ref{lf0t} shows the 12~\mic\, luminosity functions for all the
galaxies in our CDFS and HDFN surveys in the nine redshift bins
mentioned above (including the local function, plotted with a smaller
scale than in Figure~\ref{locallf}), and the fits to the data. These
fits illustrate one consequence of our choice of the Schechter
function: the contribution at high luminosities is minimized compared
with the more plausible double power-law fits.  Thus, the SCHLF fits
most probably underestimate the overall infrared output from high
luminosity galaxies, while the RUSHLF and OWNLF approaches seem to
provide better fits in this regime. A more subtle effect is that the
slope toward low luminosities is not well constrained, although it
seems to remain at a rather flat value
($-1.2\gtrsim\alpha\gtrsim-1.0$) according to the SCHLF fits and in
good agreement with the OWNLF local value. The RUSHLF case provide the
worst fits of the three sets, given that it assumes a very steep value
of $\alpha$. However, the small values of the slope may result from
incompleteness in the lowest luminosity bin fitted, plus the poor
coverage toward low luminosities in general. Thus, it is possible that
we are underestimating the number density of faint sources (i.e., the
$\alpha$ parameter). In the following Section, the effect of the large
uncertainties in the faint-end slope on the estimation of the total
luminosity density will be investigated by comparing the two extreme
cases: 1) $\alpha\sim-1$ for the SCHLF or OWNLF fits, either of which
must provide a lower limit to the total output of the population of
galaxies in each redshift bin; and 2) $\alpha\sim-1.7$ in the RUSHLF
case, which must provide an upper limit for the luminosity density,
with an important contribution from low luminosity sources.

The most robust result of the SCHLF fits is the steady growth of $L^*$
with increasing redshift. This is shown in Figure~\ref{evol} (red
stars), jointly with the evolution of the other luminosity function
parameters for each set of fits. The normalization parameter,
$\phi^*$, grows slowly (or perhaps it is nearly constant, see the
discussion below) to $z\sim0.8$, and then it seems to fall. This
observed decrease of $\phi^*$ at $z\gtrsim1$, which seems to be also
found by other authors using Schechter fittings to luminosity
functions of star-forming galaxies built at other wavelengths
(cf. gray crosses in Figure~\ref{evol}), may be strongly influenced by
the poor coverage at low luminosities, so it should be regarded with
caution. We will come back to this issue later. The fitting errors in
the slope parameter are large, and even without appealing to
incompleteness, values of $\alpha\sim-1.2$ are consistent with the
data (and compatible with other estimations of the luminosity function
of star-forming galaxies, cf. gray crosses in
Figure~\ref{evol}). Since the Schechter function is also expected to
underestimate the number of high-luminosity infrared galaxies, the
parametric fits are generally consistent with the forms of the
luminosity function found locally.

\slugcomment{Please, plot this figure with the width of two columns}
\placefigure{lf0t}

\placetable{lfpars_sch}
\begin{deluxetable*}{lcccc}
\tabletypesize{\small}
\tablewidth{380pt}
\tablecaption{\label{lfpars_sch}Results of the Schechter (1976) fits (SCHLF) to the 12~\mic\, luminosity functions.}
\tablehead{ & \colhead{$\alpha_{12}$} & \colhead{$\log(L^*_{12})$}  & \colhead{$\log(\phi^*_{12})$} & \colhead{$\log(\rho_{L_{12}})$}\\
\colhead{Redshift range} &  & \colhead{$[L_\odot]$}  & \colhead{$[\mathrm{Mpc}^{-3}\,\mathrm{Mag}^{-1}]$} & \colhead{$[L_\odot\mathrm{Mpc}^{-3}]$}}
\startdata
$0.0<z\leq0.2$ & $-1.23\pm 0.07$ &  \phn$9.61\pm0.14$ & $-2.31\pm0.16$ & $7.38\pm0.06$\\
$0.2<z\leq0.4$ & $-1.17\pm 0.08$ &  \phn$9.81\pm0.05$ & $-2.18\pm0.07$ & $7.68\pm0.02$\\
$0.4<z\leq0.6$ & $-1.05\pm 0.14$ &     $10.05\pm0.06$ & $-2.09\pm0.08$ & $7.98\pm0.03$\\
$0.6<z\leq0.8$ & $-1.03\pm 0.12$ &     $10.17\pm0.04$ & $-2.10\pm0.05$ & $8.08\pm0.02$\\
$0.8<z\leq1.0$ & $-1.06\pm 0.46$ &     $10.52\pm0.11$ & $-2.37\pm0.17$ & $8.17\pm0.23$\\
$1.0<z\leq1.4$ & $-1.16\pm 0.15$ &     $10.73\pm0.05$ & $-2.50\pm0.08$ & $8.28\pm0.03$\\
$1.4<z\leq1.8$ & $-1.15\pm 0.34$ &     $11.04\pm0.10$ & $-2.85\pm0.09$ & $8.24\pm0.04$\\
$1.8<z\leq2.2$ & $-1.05\pm 0.55$ &     $11.54\pm0.25$ & $-3.38\pm0.20$ & $8.17\pm0.12$\\
$2.2<z\leq2.6$ & $-1.05\pm 0.34$ &     $11.86\pm0.25$ & $-3.58\pm0.18$ & $8.29\pm0.11$\\
\enddata
\end{deluxetable*}

\placetable{lfpars_pow}
\begin{deluxetable*}{llccccccc}
\tabletypesize{\scriptsize}
\tablewidth{500pt}
\tablecaption{\label{lfpars_pow}Results of the fits to an evolved local luminosity function of the 12~\mic\, luminosity functions.}
\tablehead{& & \multicolumn{3}{c}{RUSHLF$^1$} & & \multicolumn{3}{c}{OWNLF$^2$}\\
\cline{3-5} \cline{7-9} \\
 & \colhead{Evol.$^3$} 
& \colhead{$\log(L^*_{12})$}  & \colhead{$\log(\phi^*_{12})$} & \colhead{$\log(\rho_{L_{12}})$}&
& \colhead{$\log(L^*_{12})$}  & \colhead{$\log(\phi^*_{12})$} & \colhead{$\log(\rho_{L_{12}})$} \\
\colhead{Redshift range}
&  & \colhead{$[L_\odot]$}  & \colhead{$[\mathrm{Mpc}^{-3}\,\mathrm{Mag^{-1}}]$} & \colhead{$[L_\odot\mathrm{Mpc}^{-3}]$}
&  & \colhead{$[L_\odot]$}  & \colhead{$[\mathrm{Mpc}^{-3}\,\mathrm{Mag^{-1}}]$} & \colhead{$[L_\odot\mathrm{Mpc}^{-3}]$} \\}
\startdata
$0.0<z\leq0.2$ &  $L$   & \phn$9.90\pm0.05$ & $-3.10\pm 0.07$ & $7.34\pm0.06$ &  & \phn$9.33\pm0.09$ & $-2.19\pm 0.10$ & $7.36\pm0.06$ \\
               & $L+D$  & \phn$9.90\pm0.05$ & $-3.10\pm 0.07$ & $7.34\pm0.06$ &  & \phn$9.33\pm0.09$ & $-2.19\pm 0.10$ & $7.36\pm0.06$ \\
$0.2<z\leq0.4$ &  $L$   &    $10.24\pm0.07$ & $-3.10\pm 0.07$ & $7.66\pm0.03$ &  & \phn$9.59\pm0.10$ & $-2.19\pm 0.10$ & $7.57\pm0.02$ \\
               & $L+D$  &    $10.09\pm0.06$ & $-2.95\pm 0.09$ & $7.65\pm0.02$ &  & \phn$9.55\pm0.10$ & $-2.14\pm 0.10$ & $7.62\pm0.02$ \\
$0.4<z\leq0.6$ &  $L$   &    $10.54\pm0.12$ & $-3.10\pm 0.07$ & $8.04\pm0.05$ &  & \phn$9.81\pm0.12$ & $-2.19\pm 0.10$ & $7.87\pm0.03$ \\
               & $L+D$  &    $10.25\pm0.09$ & $-2.82\pm 0.12$ & $8.03\pm0.04$ &  & \phn$9.73\pm0.11$ & $-2.11\pm 0.11$ & $7.85\pm0.03$ \\
$0.6<z\leq0.8$ &  $L$   &    $10.79\pm0.16$ & $-3.10\pm 0.07$ & $8.21\pm0.06$ &  &    $10.01\pm0.14$ & $-2.19\pm 0.10$ & $8.02\pm0.03$ \\
               & $L+D$  &    $10.39\pm0.12$ & $-2.70\pm 0.15$ & $8.21\pm0.04$ &  & \phn$9.90\pm0.13$ & $-2.07\pm 0.13$ & $8.04\pm0.02$ \\
$0.8<z\leq1.0$ &  $L$   &    $10.85\pm0.20$ & $-3.10\pm 0.07$ & $8.29\pm0.24$ &  &    $10.11\pm0.17$ & $-2.19\pm 0.10$ & $8.15\pm0.24$ \\
               & $L+D$  &    $10.61\pm0.15$ & $-2.84\pm 0.18$ & $8.31\pm0.24$ &  &    $10.16\pm0.15$ & $-2.24\pm 0.14$ & $8.15\pm0.23$ \\
$1.0<z\leq1.4$ &  $L$   &    $10.93\pm0.25$ & $-3.10\pm 0.07$ & $8.36\pm0.10$ &  &    $10.23\pm0.20$ & $-2.19\pm 0.10$ & $8.26\pm0.06$ \\
               & $L+D$  &    $10.74\pm0.19$ & $-2.86\pm 0.22$ & $8.42\pm0.07$ &  &    $10.35\pm0.17$ & $-2.37\pm 0.16$ & $8.20\pm0.04$ \\
$1.4<z\leq1.8$ &  $L$   &    $11.02\pm0.30$ & $-3.10\pm 0.07$ & $8.46\pm0.12$ &  &    $10.38\pm0.24$ & $-2.19\pm 0.10$ & $8.41\pm0.07$ \\
               & $L+D$  &    $11.28\pm0.23$ & $-3.47\pm 0.27$ & $8.35\pm0.08$ &  &    $10.82\pm0.21$ & $-2.86\pm 0.18$ & $8.18\pm0.04$ \\
$1.8<z\leq2.2$ &  $L$   &    $11.17\pm0.35$ & $-3.10\pm 0.07$ & $8.60\pm0.18$ &  &    $10.55\pm0.28$ & $-2.19\pm 0.10$ & $8.58\pm0.14$ \\
               & $L+D$  &    $12.08\pm0.27$ & $-4.23\pm 0.31$ & $8.38\pm0.15$ &  &    $11.33\pm0.24$ & $-3.36\pm 0.20$ & $8.19\pm0.12$ \\
$2.2<z\leq2.6$ &  $L$   &    $11.42\pm0.39$ & $-3.10\pm 0.07$ & $8.86\pm0.18$ &  &    $10.80\pm0.31$ & $-2.19\pm 0.10$ & $8.83\pm0.13$ \\
               & $L+D$  &    $12.21\pm0.30$ & $-4.18\pm 0.35$ & $8.56\pm0.15$ &  &    $11.63\pm0.26$ & $-3.46\pm 0.22$ & $8.39\pm0.11$ \\
\enddata
\tablecomments{$^1$ For the RUSHLF fits, the slopes at low and high 
luminosities are fixed: $\alpha_{12}=-1.70\pm0.02$ and
$\beta_{12}=-3.60\pm0.09$.  $^2$ For the OWNLF fits, the slopes are
also fixed: $\alpha_{12}=-1.17\pm0.07$ and $\beta_{12}=-2.97\pm0.16$.
$^3$The type of evolution can be: luminosity evolution ($L$), or a
combined luminosity plus number density evolution ($L+D$).}
\end{deluxetable*}

For the RUSHLF and OWNLF fits, the growth of $L^*$ with redshift is
again a robust result, as shown by the green and blue stars in the
first panel of Figure~\ref{evol} \footnote{Note that there is an
offset between the values of the $L^*$ and $\phi^*$ parameters for the
three sets of fits carried out in the paper. This is due to the fact
that each set of fits uses a different parametrization: a Schechter
function with an adjustable low luminosity slope or a double power-law
with a fixed low luminosity slope. With an adjustable low luminosity
slope, the fitting routine systematically settles on solutions with
smaller $L^*$ and larger $\phi^*$ values, and flatter slopes than with
the double power law. The uncertainties in the bright-end slope
$\beta$ in the double power-law parametrization are also correlated
with the uncertainties in $\phi^*$ and $L^*$. This explains the
offsets between the results for the RUSHLF and OWNLF fits.}. This
result is obtained for both pure luminosity evolution (open stars) and
combined luminosity plus density evolution (filled stars), although
the rate of change varies from one type of evolution to the other (and
from these to the SCHLF approach). The rate of change from $z=0$ to
$z\sim1$ seems to follow a $(1+z)^{n_L}$ law, where $n_L=3-5$. In the
region of overlap ($z\lesssim1.1$), this behavior agrees closely with
the results of \citet{emeric05}.  The uniform pure density evolution
seems to be ruled out by our data, given that it fails to describe the
luminosity function points as early as $z\sim0.4$. Indeed, the
probability of exceeding by chance the $\chi^2$ value obtained for the
fit with a pure density evolution is 7\% for the RUSHLF fits (5\% for
OWNLF). In contrast, the probability of exceeding by chance the
$\chi^2$ value obtained for the fit with pure $L$ or $L+D$ evolution
is 75\% (80\%) and 90\% (93\%), respectively. This means that we are
not able to confidently break the degeneracy between these two
scenarios, although the fits for the combined $L+D$ evolution are
slightly better. That is, the luminosity function data points can be
reproduced with a strong $L$ evolution ($n_L\sim4-5$), or with a
weaker luminosity evolution ($n_L=2.6-3.1$) combined with a relatively
smaller density evolution ($n_D=0.5-2.0$). When we used Schechter
functions, which also allowed changes in the faint end slope, the
combined $L+D$ evolution is also slightly favored.

Figure~\ref{evol} also shows the best linear fits to the evolution of
the luminosity function parameters at $0<z<0.8$. In the case of the
Schechter fitting, the number density of sources, as parameterized by
$\phi^*$, evolves as $(1+z)^{0.9\pm0.6}$ in this redshift range. The
typical infrared luminosity ($L^*$) evolves as $(1+z)^{3.1\pm0.5}$. In
the OWNLF fits, the $L+D$ evolution predicts
$L^*\propto(1+z)^{3.0\pm0.3}$ and $\phi^*\propto(1+z)^{0.6\pm0.2}$
(green dashed lines in the first two panels). In comparison, when we
fit the data with a pure $L$ evolution in the OWNLF case, $\phi^*$
remains obviously constant, and $L^*$ behaves similarly to the
Schechter function fits, evolving as $(1+z)^{3.6\pm0.3}$ (green dotted
lines in the first two panels). As we mentioned before, the RUSHLF
fits are always considerably worse than the OWNLF and SCHLF ones, but
they predict similar evolution laws: $L^*\propto(1+z)^{2.6\pm1.1}$ and
$\phi^*\propto(1+z)^{2.1\pm0.6}$ for the $L+D$ case, and
$L^*\propto(1+z)^{4.7\pm0.3}$ for the pure $L$ evolution. Combining
the results from the three sets of fits, the most probable values of
the exponents of the evolution laws are: $n_L=3.0\pm0.3$ and
$n_D=1.0\pm0.3$.

Above $z\sim0.8$, the evolution degeneracy is more severe. The
Schechter fits seem to favor a scenario where $\phi^*$ decreases
steadily as $(1+z)^{-5.1+0.8}$, and $L^*$ continues rising as
$(1+z)^{4.8\pm0.8}$, i.e., a few galaxies with very violent star
formation dominate the TIR luminosity density. However, our data could
also be reproduced with a slower (or even null) decrease of $\phi^*$
up to $z\sim3$ and an also slower increase of $L^*$ (see open stars in
the left two panels of Figure~\ref{evol}).

\slugcomment{Please, plot this figure with the width of two columns}
\placefigure{evol}

\subsection{Cosmic star formation rate density}
\label{sfrd}

\begin{figure*}
\begin{center}
\includegraphics[angle=-90,width=12cm]{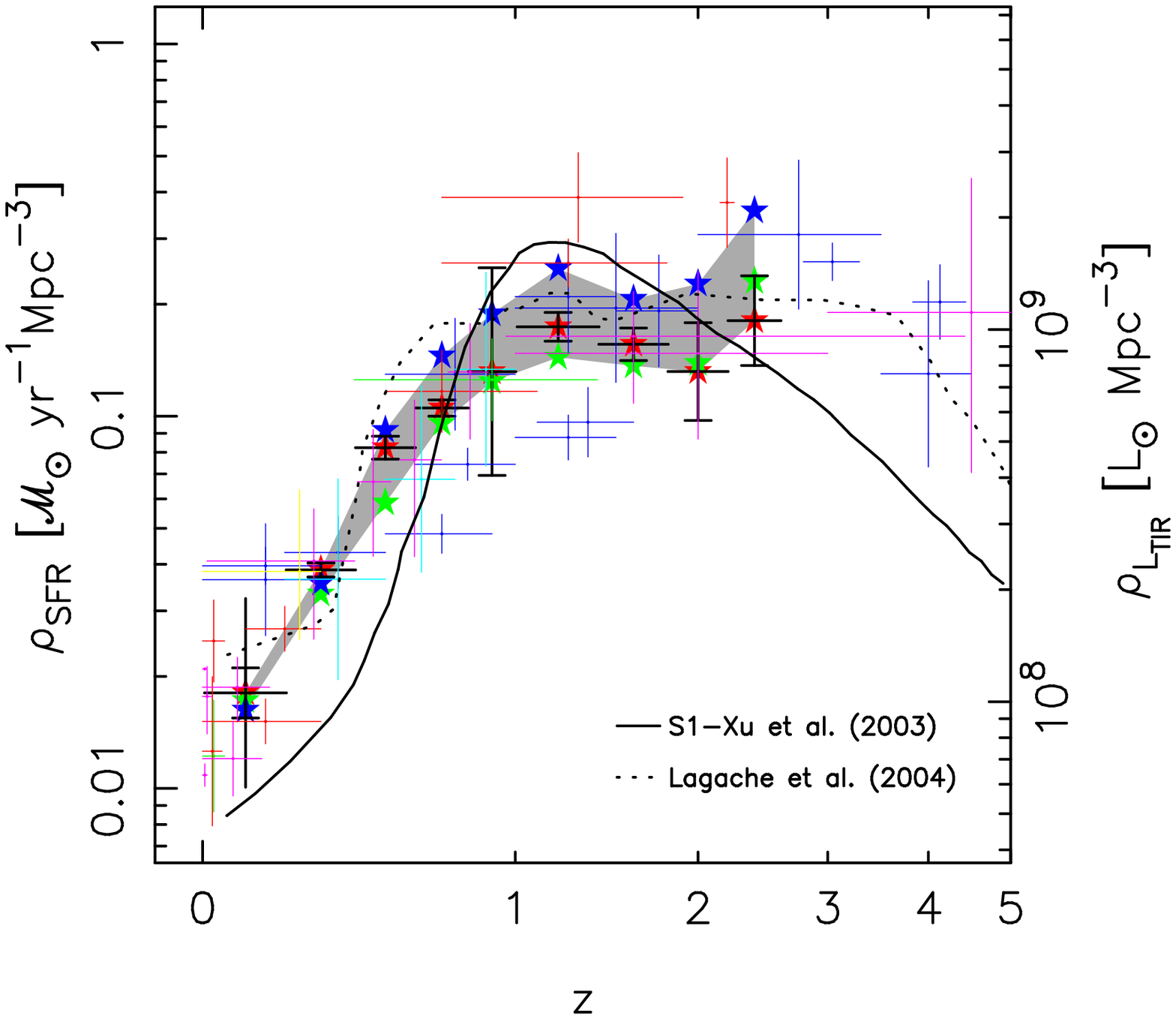}
\figcaption{\label{mad}The Lilly-Madau
diagram \citep{1995ApJ...455..108L,1996MNRAS.283.1388M}: evolution of
the SFR density of the Universe with redshift. The estimations based
on the SCHLF, RUSHLF, and OWNLF fits are plotted with red, blue, and
green stars, respectively (see text for details). The shaded area
delimits the zone between the two extreme SFR density estimations for
each redshift. The heavy error bar at $z\sim0.1$ shows the uncertainty
in the transformation from the monochromatic 12~\mic\, luminosity to
the TIR emission, as shown in Equation~\ref{12tofir} (see the text for
a discussion of this error). This error is common for all our points,
but it is only given in the first one for clarity. Vertical segments
for each point show the uncertainty related to the integration of the
SCHLF luminosity function (comparable to the OWNLF and RUSHLF
cases). The horizontal lines show the range of redshifts used in each
bin. The curves show two typical models: one with a decay from
$z\sim1$ \citep[][S1 model]{2003ApJ...587...90X}, and another with a
constant SFR density at high redshift
\citep{2004ApJS..154..112L}. The colored points (shown with error bars) 
are extracted from different sources in the literature, normalized to
the same cosmology by \citet{2004ApJ...615..209H}. Red symbols are
estimations based on H$\alpha$ or H$\beta$ measurements
\citep{1995ApJ...455L...1G,1998ApJ...508..539P,1998ApJ...495..691T,
1999MNRAS.306..843G,1999ApJ...519L..47Y,2000A&A...362....9M,
2000AJ....120.2843H,2000MNRAS.312..442S,2002MNRAS.337..369T,2003ApJ...591..827P}. Green
symbols stands for $[OII]\lambda3737$ estimations
\citep{1997ApJ...481...49H,1998ApJ...504..622H,2002ApJ...570L...1G,
2003ApJ...589..704T}. UV-based data points are plotted in blue
\citep{1996ApJ...460L...1L,1997ApJ...486L..11C,1998MNRAS.300..303T,
1999AJ....118..603C,1999ApJ...519....1S,2000MNRAS.312..442S,
2001ApJ...559L.105M,2002AJ....124.1258W,2004ApJ...600L.103G}. Cyan
estimations are based on mid-infrared data
\citep{1999ApJ...517..148F}. Magenta points are based on sub-mm and radio 
observations \citep{1989ApJ...338...13C,1998Natur.394..241H,
2000AJ....119.2092B,2000A&A...360..463M,2000ApJ...544..641H,2002MNRAS.329..227S,
2002MNRAS.330..621S,2002AJ....124..675C}. The yellow point is based on
X-ray data \citep{2003MNRAS.345..939G}.}
\end{center}
\end{figure*}

In the previous Section, we presented two ways of fitting the
luminosity functions: one using a Schechter function to fit the data
independently at each individual redshift range, and the other
assuming a constant shape of the luminosity function (the local shape,
which is the best constrained) and evolving it with redshift in
density or/and luminosity. The former technique allows a change in the
shape of the luminosity function, mostly in the faint end slope, while
the latter procedure fixes this slope. It also assumes a less rapid
fall at high-luminosities, which seems to be the case in the local
Universe. In this Section, we integrate the SCHLF, OWNLF, and RUSHLF
fits of the luminosity functions to get different (possibly biased)
estimates of the total 12~\mic\, luminosity density of the Universe at
$0<z\lesssim3$, the TIR luminosity density, and the SFR density. The
TIR luminosities have been estimated using
Equation~\ref{12tofir}. These luminosity densities will be translated
to cosmic SFR densities in Figure~\ref{mad} using
Equation~\ref{fir2sfr}. At the end of this Section, by comparing the
results obtained with the three sets of fits, we will discuss how
changes in the faint end and bright end slopes affect the estimations
of the luminosity and SFR densities.

\slugcomment{Please, plot this figure with the width of two columns}
\placefigure{mad}

Figure~\ref{mad} shows the evolution of the SFR density of the
Universe as a function of redshift (red, green, and blue stars
referring to the integration of the SCHLF, OWNLF, and RUSHLF fits,
respectively) in a Lilly-Madau diagram
\citep{1995ApJ...455..108L,1996MNRAS.283.1388M}. Our survey reproduces
the rapid increase in $\rho_\mathrm{SFR}$ from $z=0$ to $z\sim1.4$
observed by many previous works. Our estimations follow a
$(1+z)^{4.0\pm0.2}$ law up to $z=0.8$ (i.e., $\beta=4.0\pm0.2$), and a
lower slope ($\beta\sim3.4$) up to $z\sim1.4$. For $z<1$, we obtain:

\begin{equation}
\log{(\rho_\mathrm{SFR})}=(-1.87\pm0.04)+(3.98\pm0.22)\times\log{(1+z)}
\end{equation}

This result is consistent with that of \citet[][see also
\citealt{astro-ph/0105280}]{2004ApJ...615..209H}, who used all the 
SFR density estimations plotted in Figure~\ref{mad} (obtained with
different SFR tracers), although our slope is higher (and our local
density is slightly smaller): \citet{2004ApJ...615..209H} gives
$\beta=3.10\pm0.25$ applying a simple obscuration correction for the
SFR density estimations, and $\beta=3.29\pm0.26$ for a
luminosity-dependent obscuration correction. The analysis of
\citet{2004ApJ...615..209H} is partly based on UV surveys, 
which tend to obtain significantly less steep values of the SFR
density evolution slope ($\beta=2-2.5$, see
\citealt{2005ApJ...619L..47S,2005MNRAS.tmp..165B}). However, other 
works based on UV and optical surveys (using different emission-lines)
find larger values, closer to our estimation (e.g.,
\citealt{1996ApJ...460L...1L,2002MNRAS.337..369T,2003A&A...402...65H}).
Moreover, models based on IR and sub-millimeter models also predict an
evolution with an exponent close to $\beta\sim4$
\citep{1999MNRAS.302..632B,2003ApJ...587...90X,2004ApJS..154..112L}. 

The scatter in the results for the slope of the evolution of
$\rho_\mathrm{SFR}$ suggests that the extinction properties of the
galaxies dominating the total SFR density are evolving with
redshift. For example, our estimation of the SFR density at $z\sim0.1$
(which is consistent with other estimations based on radio
observations, e.g., \citealt{1989ApJ...338...13C,2002MNRAS.329..227S,
2002MNRAS.330..621S}) is $\sim$40\% lower than the most recent results
achieved by $H\alpha$
\citep{2003ApJ...591..827P} or UV/optical surveys \citep{2003ApJ...587...55G,
2005ApJ...619L..15W, 2005ApJ...619L..47S,2005ApJ...619L..59M} in the
local Universe. This may indicate that the star formation in this
redshift regime is dominated by galaxies with not very extincted
bursts, where the dust emission only traces a small part of the total
SFR of each galaxy (i.e., many photons from the newly-born stars do
not interact with the dust, but they escape through the UV or
emission-lines). This effect would be reasonable in galaxies with not
very intense star formation
($\mathrm{SFR}\lesssim5\,\mathcal{M}_\odot\,\mathrm{yr}^{-1}$), given
that the most violent star-forming galaxies show the highest dust
attenuations \citep{2001AJ....122..288H,2001ApJ...558...72S,
2003ApJ...591..827P}. Those low intensity star-forming galaxies
contribute importantly to the total SFR density in the local Universe
\citep[see Figure 16 in ][]{2003MNRAS.338..525P}. As we move to higher
redshifts, the star formation in galaxies starts to be dominated by
intense dust-enshrouded bursts, whose SFR is better traced by the TIR
emission \citep{2003ApJ...584...76C}. In this scenario, the evolution
of the extinction properties also seems to be consistent with the
increasing contribution of IR-bright galaxies with dust enshrouded
bursts to the cosmic SFR density (\citealt{2001ApJ...556..562C}; see
also Figure~\ref{evollum}).

After the increase from $z=0$ to $z\sim1.4$, we find a roughly flat
behavior of $\rho_\mathrm{SFR}$ up to $z\sim3$, very similar to what
some models predict (e.g., \citealt{2004ApJS..154..112L}, shown in the
figure, or \citealt{2001ApJ...556..562C}), and consistent with the
results from most UV/optical surveys \citep[see][and references
therein]{2004ApJ...615..209H}.

There are three issues affecting our estimations in the Lilly-Madau
diagram: how the fitting procedures affect the results, the
translation of 12~$\mu$m luminosity to bolometric infrared luminosity,
and of bolometric infrared luminosity to true bolometric luminosity
and star formation rates.

For the first of the issues, by integrating under the luminosity
functions of a given shape, we found that the TIR output of the galaxy
population was virtually independent of whether the evolution was in
luminosity, density, or a mixture (within the uncertainties; see
Table~\ref{lfpars_pow}). By varying the shape of the fitting curve, we
also found that the total output of the galaxy population was not
strongly dependent on the shape of the luminosity function above
$L^*$. This is shown with green stars in Figure~\ref{mad}. These
points were calculated using the OWNLF fits, i.e, an almost flat
behavior at the faint end and a less steep behavior than the Schechter
function at the bright end. It appears that the luminosity function is
too steep in the high-luminosity region for plausible variations to
change the integral of the luminosity function significantly (average
change smaller than 20\%). Given that AGNs should predominantly
populate the bright end of the luminosity function, it seems probable
that they do not affect our results by a large factor,
either\footnote{For example, the AGNs removed from our sample as
discussed in Section~\ref{agns}.}.

For a fixed behavior toward high luminosity, we found a change by a
factor of two as $\alpha$ was changed from $-1.7$ (as in the RUSHLF
case and some UV surveys) to approximately $-1.0$ (as the SCHLF and
OWNLF estimations predict). The range of estimates is shown in
Figure~\ref{mad} with the shaded area. Therefore, the uncertainties in
the TIR luminosity density include a significant contribution from the
lack of knowledge of the low luminosity galaxy population. At low
redshift, $\alpha$ seems to be close to the flat value:
$-1.3\lesssim\alpha\lesssim-1.0$ (\citealt{2002MNRAS.330..621S},
\citealt{2003ApJ...591..827P}, \citealt{2005ApJ...619L..15W}, 
\citealt{2005ApJ...619L..31B}). The most
recent estimations of $\alpha$ based on the deepest observations of
high-redshift galaxies emitting strongly in the UV
\citep{2004A&A...421...41G} suggest also an almost flat value with a
marginal indication of evolution of the slope with redshift (to
shallower values). This means that the SFR densities should be closer
to the SCHLF or OWNLF values than to the upper limits of the RUSHLF
case (quoted with the blue stars in Figure~\ref{mad}).


The very weak dependence of the integral of the luminosity function on
either very high or very low (for $\alpha\sim-1$) luminosity galaxies
also means that photometric redshift outliers have little effect on
this integral. In addition, the integral of the luminosity function is
always a more robust calculation than the individual parameters of the
fitting function. The Monte Carlo simulation described in
Appendix~\ref{montecarlo} confirms these statements.

The second of the issues affecting our SFR density estimates is
related to the uncertainties in the monochromatic-to-TIR emission
relationship. This error is roughly a factor of two for individual
galaxies (based on Equation~\ref{12tofir}, extracted from
\citealt{2001ApJ...556..562C}). We show this range on the first point 
(at $z\sim0.1$) in Figure~\ref{mad}. Fortunately, these uncertainties
can be reduced by future work including the longer wavelength MIPS
bands and ground-based data in the sub-millimeter. Moreover, for
calculating the TIR luminosity density, we average the luminosities of
galaxies in luminosity bins, and integrate them to all the possible
values. In this averaging and integration procedure, it is probable
that the final uncertainties diminish.

The third issue in the SFR density estimation appears to be less of a
problem as we move to higher redshifts. The unaccounted contribution
of ultraviolet luminosity is probably lower than $40\%$ of the
infrared contribution for all redshifts. In this regard,
\citet{astro-ph/0502246} quote that the UV luminosity density is
$\sim7$ times smaller than the IR density at $z\sim0.7$. In
\citet{2005ApJ...619L..63B}, an average UV attenuation of
$A_\mathrm{FUV}=2.7$ for a sample of ULIRGs at $z<1.6$ is
found. Attenuations of the same order (factors of $5-10$) have also
been found by other authors \citep[][among
others]{1998ApJ...498..106M,1999ApJ...519....1S,2000ApJ...544..218A,
2001ApJ...559L.105M,2005ApJ...619L..47S}. This means that the star
formation traced by the UV alone (without extinction correction, i.e.,
the star formation which the IR cannot trace because it did not heat
the dust) is $5-10$ times smaller than the star formation traced by
the IR. If there is an evolution in the extinction properties of
galaxies, as we previously discussed, these uncertainties will yield
an offset of the total SFR density which depends on the redshift. If
the evolution is not present, there should be a systematic offset of
the estimates (i.e., all the points in the Lilly-Madau plot would move
up approximately by the same quantity). A comparison of the extinction
properties of the galaxies selected in UV/optical and IR surveys will
be necessary to address this issue.

\subsection{Contribution of galaxies with different TIR luminosities and masses 
to the total SFR density of the Universe}
\label{fin}

\begin{figure}
\begin{center}
\includegraphics[angle=-90,width=9cm]{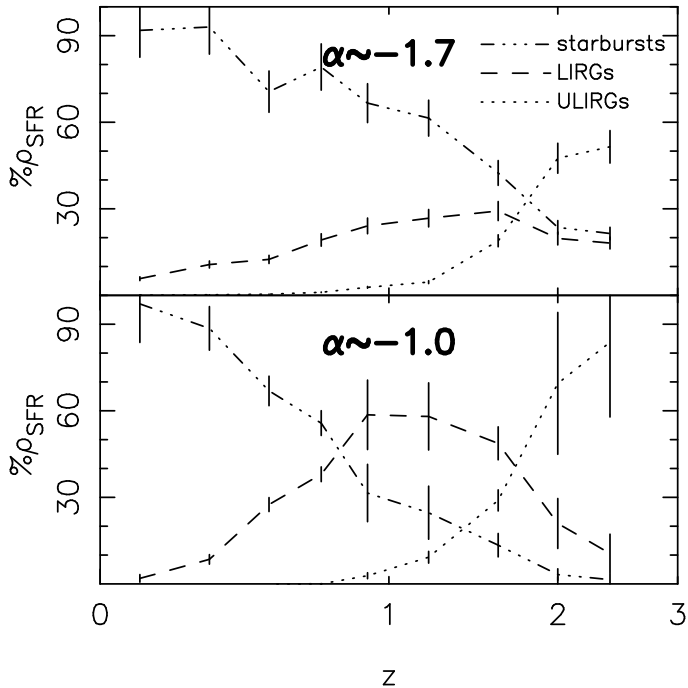}
\figcaption{\label{evollum}Relative contribution of starbursts 
($L_\mathrm{TIR}<10^{11}\,L_\sun$), LIRGs
($10^{11}<L_\mathrm{TIR}<10^{12}\,L_\sun$), and ULIRGs
($L_\mathrm{TIR}>10^{12}\,L_\sun$) to the total SFR density of the
Universe as a function of redshift. The two extreme cases of the
faint-end slope value are separated: $\alpha\sim-1.7$ (RUSHLF fits,
upper panel) and $\alpha\sim-1.0$ (SCHLF case, lower panel). The error
bars show the uncertainties on the integration in each luminosity
range, considering the errors of the individual luminosity function
parameters.}
\end{center}
\end{figure}

\slugcomment{Please, plot this figure with the width of one column}
\placefigure{evollum}

Figure~\ref{evollum} shows the contribution of galaxies of different
TIR luminosities to the integrated SFR (or TIR luminosity) density of
the Universe as a function of redshift. We have produced two plots to
account for the two extreme cases of the faint-end slope value:
$\alpha\sim-1.0$ (SCHLF case) and $\alpha\sim-1.70$ (RUSHLF fits). In
the almost flat luminosity function scenario, there is a dominant but
decreasing contribution of normal and starburst galaxies with faint
infrared luminosities ($L_\mathrm{TIR}<10^{11}\,L_\sun$) to the total
SFR density up to at least $z\sim0.8$. At this redshift, LIRGs already
form approximately half of the total amount of newly-born stars. These
results are consistent with the ones achieved by
\citet{emeric05} using CDFS data and photometric redshifts from COMBO17. 
The evolution of LIRGs was expected by ISO-based models, such as
\citet{2001ApJ...556..562C,2004ApJS..154...80C} and
\citet{2003ApJ...587...90X}, which predicted a $\sim70$\% contribution
to the total luminosity density for $z>0.5$. Our estimation is
somewhat below this value. The evolution of LIRGs decelerates at
$z\sim0.9$, remaining approximately at the $50-60$\% level up to
$z\sim1.5$, while starbursts continue their decline, and ULIRGs start
to contribute significantly to the total luminosity density. By
$z\sim2$, ULIRGs already form more than 70\% of the newly-born stars
in the Universe, and they completely dominate the luminosity density
at $z\sim3$.

Figure~\ref{evollum} also shows the contributions to the total SFR
density obtained with the RUSHLF case, i.e., considering a rather
steep luminosity function for all redshifts (upper panel). As shown in
Figure~\ref{mad}, this gives an upper value for the SFR density of the
Universe, with an important contribution from galaxies with modest
star formation. Indeed, the upper panel of Figure~\ref{evollum} shows
that starbursts still show a dominant but decreasing contribution to
the total SFR density at $z<0.6$, but their contribution is not
negligible at $z\sim2.0$ (they still form roughly 30\% of the total
amount of stars at that redshift). The contribution of LIRGs rises
slowly up to $\sim30$\% at $z\sim1.5$, and then stays approximately
constant up to $z\sim3.0$. Starting at $z\sim1$, ULIRGs start to
contribute non-negligibly to the total SFR density, and reach
$\sim$30\% of the total SFR density at $z\sim2$. Note that the
contribution from starbursts to the total SFR density is larger in the
flat slope case than in the steep case for $z\lesssim0.3$ (and,
consequently, the contribution of LIRGs to the total density is
smaller in the $\alpha\sim-1.0$ case). This effect is directly related
to the uncertainties in the fits (in $L^*$ and $\alpha$) at low
redshift.

\begin{figure*}
\begin{center}
\includegraphics[angle=-90,width=13cm]{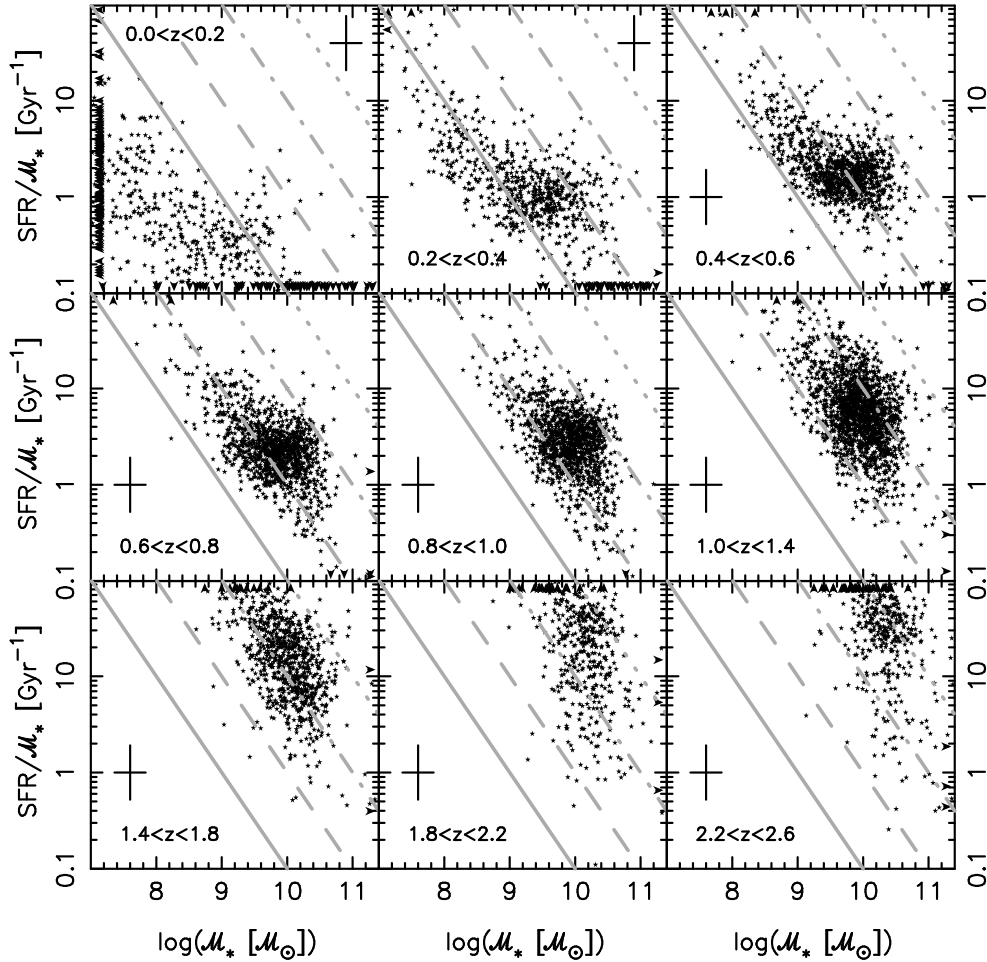}
\figcaption{\label{spec}Relationship between the specific SFR
(SFR per stellar mass unit) and the total stellar mass for the
galaxies in our 24~\mic\, survey. The sample is divided in 9 different
redshift bins, each plotted in one panel. The lines, from left to
right, correspond to constant SFRs of 1, 10, 100, and
1000~$\mathcal{M}_\odot\,\mathrm{yr}^{-1}$. Typical errors for each
axis are also shown.}
\end{center}
\end{figure*}

Figure~\ref{evollum} demonstrates the shift of the star formation
density to LIRGs and ULIRGs (i.e., to IR-bright galaxies with very
violent dust enshrouded bursts of star formation) as we move from
$z=0$ to $z\gtrsim1$. LIRGs and ULIRGs tend to be the most massive
star-forming galaxies at $z\lesssim1$, with masses
$\mathcal{M_*}\sim10^{11}\,\mathcal{M}_\odot$
\citep[see, e.g.,][]{2002ApJ...580..789R,2002ApJ...580...73T,
2003A&A...403..501F}. It is interesting to study the connection
between dust enshrouded star formation and the stellar mass of each
galaxy. We analyze the relationship between these two parameters in
Figure~\ref{spec}.

We estimated stellar masses for all our galaxies using $K$-band
luminosities calculated by interpolation among the templates utilized
to get the photometric redshifts. By using the IRAC photometry, we
could probe the rest-frame $K$-band up to $z\sim3$. For local
galaxies, a number of authors have demonstrated how accurate stellar
masses can be determined from $K$-band luminosities combined with
optical colors to constrain the star formation history
\citep{2001ApJ...550..212B,2003ApJS..149..289B,2003MNRAS.341...33K}. 
They show that the mass-to-light ratios for $K$-band luminosities
should not change more than a factor of 2 to 3 across a wide range of
star formation histories, in comparison with a factor of $>$10 for
optical mass-to-light ratios. The effects of extinction are also
negligible in the near infrared.

At higher redshifts, due to galaxy evolution, the mass-to-light ratio
should decrease \citep{2001ApJ...562L.111D,2004ApJ...608..742D,
2004A&A...424...23F}. We therefore based our mass estimates on the
redshift-dependent relationships for late-type galaxies found in
\citet{2004A&A...424...23F}. At low redshift, this procedure agrees well 
(with less than a 10\% scatter) with, for example, that of
\citet{2003ApJS..149..289B}.  However, for $0.2\lesssim z\lesssim3$,
the values are down to 2 times lower than the local relationship found
by \citet{2003ApJS..149..289B}. A lower limit to the masses can also
be determined by starburst modeling, based on the galaxy
luminosities. The models of M82
\citep{1993ApJ...412...99R,1993ApJ...412..111M,2003ApJ...599..193F} 
suggest that the star formation in the infrared-luminous galaxies
would produce a stellar population only slightly (no more than a
factor of two) lower in mass than the value determined by the method
of \citet{2004A&A...424...23F}. We conclude that our mass estimates
are accurate to a factor of 2 to 3, which is sufficient for the
following qualitative analysis of the connection between star
formation and stellar mass as a function of redshift. More robust
estimates of the stellar masses (leading to a more detailed study of
the the SFR-mass connection) should rely on the modeling of the
stellar populations in each galaxy \citep[see, e.g.,
][]{2001ApJ...559..620P,2003MNRAS.341...33K,2003MNRAS.338..508P,
2003MNRAS.338..525P}, a topic that will be addressed in future works.

\slugcomment{Please, plot this figure with the width of two columns}
\placefigure{spec}

Figure~\ref{spec} shows the relationship between the specific SFRs
(SFR per stellar mass unit) and the total stellar masses for all the
galaxies in our survey, divided into redshift bins. As we move to
higher redshifts, there is a trend for the most active star-forming
galaxies to be more massive: the cloud of points for each redshift
range shifts to larger stellar masses and specific SFRs as we move to
higher redshifts. The trend is in part due to selection effects, i.e.,
we are only detecting the IR brightest objects at the furthest
distances, and those objects should also be bright in the optical and
NIR, thus presenting large stellar masses. However, it is worth
pointing out that our survey is able to detect the galaxies that
dominate the SFR density up to $z\sim3$, and those galaxies are more
and more massive. 

This result seems to support the theory of ``downsizing''
\citep{1996AJ....112..839C}, for which there is an increasing amount
of observational evidence
\citep{2004Natur.428..625H,2004Natur.430..181G,
2005ApJ...619L.135J,2005ApJ...621L..89B}. In this theory, the most
massive galaxies form first in very violent episodes of star
formation, while the formation of less massive systems continues as we
move to lower redshifts. \citet{2005ApJ...619L.135J} argue that the
SFR in the most massive galaxies
($\mathcal{M_*}>10^{10.8}\,\mathcal{M}_\odot$) was much larger at
$z\sim2$ than in the local Universe. This is supported by
Figure~\ref{spec} because most of the galaxies with
$\mathcal{M_*}>10^{10.6-10.8}\,\mathcal{M}_\odot$ and high specific
SFRs are placed at $z>1.5$. That mass value is, in fact, a rather
sharp cut-off for the entire sample, which should have its origin in
the steep fall of the stellar mass function
\citep[see][]{2004ApJ...608..742D,2004A&A...424...23F,2005MNRAS.358L...1F},
but may also be due to the most massive systems having formed the bulk
of their stars in an epoch before $z=1.5-2.5$ in very violent episodes
of star formation, probably presenting very high IR
luminosities. Beyond that point, their specific SFRs decrease
considerably, gradually disappearing from our plot. This scenario
would also be supported by the results about the population of
high-mass galaxies at $z=1-3$, that require ULIRGs at $z>3$
\citep[see, e.g., ][]{2004ApJ...614L...9M,2005astro.ph..4219L}. 
For intermediate-mass galaxies
($10^{10.0}\lesssim\mathcal{M_*}\lesssim10^{10.6}\,\mathcal{M}_\odot$)
with high SFRs, the observed redshift distribution peaks in the range
$1.0\lesssim z\lesssim2.5$, also in agreement with
\citet{2005ApJ...619L.135J}. The dominant sources in our survey at 
redshifts $z\lesssim1$, just when the cosmic SFR density starts its
decline, present $\mathcal{M_*}\lesssim10^{10}\,\mathcal{M}_\odot$ and
$\mathrm{SFR}/\mathcal{M_*}\lesssim8\,\mathrm{Gyr}^{-1}$.


\section{CONCLUSIONS}
\label{conclu}

We have cross-correlated the sources detected by MIPS at 24~\mic\, in
the Chandra Deep Field South and Hubble Deep Field North with
ultraviolet, optical, near-infrared, and mid-infrared (IRAC) catalogs.

Using this multiwavelength dataset, we have estimated the photometric
redshifts of all the sources in our sample. The technique used to
estimate these redshifts is based on empirically-built templates
obtained from sources with know spectroscopic redshifts. The accuracy
of these redshifts is better than 10\% for 80\% of the sample. As a
test of our conclusions based on this redshift estimation, we show
that our results for $0<z\lesssim1$ closely agree with those in a
companion paper by \citet{emeric05}. The derived redshift distribution
of the sources detected by our survey (for fluxes $F_{24}>83$~$\mu$Jy)
peaks at around $z=0.6-1.0$, and decays monotonically from $z\sim1$ to
$z\sim3$. The shape of this decay is not reproduced by existing models
of galaxy evolution.

We have also obtained mid-infrared monochromatic and total infrared
luminosities for all the sources, and have built luminosity functions
at 12~\mic. According to our results, the local luminosity function is
relatively flat at faint luminosities ($\alpha\sim-1.2$). Given the
limitations in our data (in detection limit and areal coverage), we
estimated the luminosity functions in a number of ways that allowed us
to understand the systematic errors. By fitting
\citet{1976ApJ...203..297S} functions and forms of the local
luminosity function to the luminosity function data points in
different redshift bins, we find: 1) the normalization of the
luminosity function, $\phi^*$, could be flat or increase at a maximum
rate of $(1+z)^{2.1\pm0.6}$ up to $z\sim0.8$; at higher redshifts, it
seems to stay constant or decrease with redshift; 2) the typical
luminosity, $L^*$, increases as at least $(1+z)^{2.6\pm1.1}$ to
$z\sim0.8$, and continues to increase at a roughly similar rate to
higher $z$; 3) the best fits to our data predict an evolution where
$L^*\propto(1+z)^{3.0\pm0.3}$ and $\phi^*\propto(1+z)^{1.0\pm0.3}$;
and 4) the low luminosity slope, $\alpha$, is not well constrained
between the values of $\alpha\sim-1.0$ to $\alpha\sim-1.7$, but the
best fits to our data indicate an almost flat value
$\alpha\lesssim-1.3$.

We reproduce the previously seen rapid increase of the total infrared
luminosity density of the Universe (and the cosmic star formation rate
density) up to $z\sim1.4$. This increase follows a $(1+z)^{4.0\pm0.2}$
law up to $z=0.8$, with a declining rate up to $z\sim1.4$. The slope
at $z<1$ is lower than that observed by some UV surveys, possibly
indicating an evolution in the extinction properties of galaxies. At
$z>1.4$, we find no evidence for a decrease in the SFR density, but
find a flat distribution up to $z\sim3$. Uncertainties in the faint
end slope of the luminosity functions could affect these results
significantly.

Assuming an almost flat slope at faint luminosities for the luminosity
functions at $z>0$, our results indicate that the SFR density is
dominated at low redshift ($z\lesssim0.5$) by galaxies which are not
very luminous in the infrared ($L_\mathrm{TIR}<10^{11}\,L_\sun$). The
contribution from luminous infrared galaxies
($10^{11}<L_\mathrm{TIR}<10^{12}\,L_\sun$) increases rapidly from
$z\sim0.4$, forming approximately half of the total amount of
newly-born stars by $z\sim0.7$, while the starburst population
declines steadily. At $z=1$, Ultraluminous Infrared Galaxies
($L_\mathrm{TIR}>10^{12}\,L_\sun$) start to play a role, probably
dominating the cosmic SFR density at $z\gtrsim2$. If we consider
steeper values of the slope at faint luminosities for the luminosity
functions at $z>0$, the contribution to the total SFR density of
starbursts is larger at high $z$, the evolution of LIRGs (relative to
starbursts) is not as marked as in the case of a flat slope, and all
three galaxy types (starbursts, LIRGs, and ULIRGs) form approximately
the same amount of stars at $z\sim2.5$. The rapid increase of $L^*$
with $z$ and our division of the cosmic star formation rate density
according to the luminosities of the contributing galaxies both
agree. The role of ULIRGs in the overall star formation increases
rapidly for $z\gtrsim1.3$.

Finally, the distribution of masses and specific SFRs (SFR per stellar
mass unit) of the galaxies in our survey seems to support a
``downsizing'' galaxy formation scenario, where the most massive
galaxies would form first ($z\gtrsim2$), and the less massive systems
would be continuously forming down to lower redshifts.

\acknowledgments

Support for this work was provided by NASA through Contract Number
960785 issued by JPL/Caltech.  This work is based on observations made
with the \spitzer\, Space Telescope, which is operated by the Jet
Propulsion Laboratory, California Institute of Technology under NASA
contract 1407. We thank an anonymous referee for her/his very
constructive comments. We would like to thank Jim Cadien for his very
useful work in data reduction and analysis. We are very grateful to
Dr. T.~A. Small for his help with the GALEX data. We are also grateful
to Dr. R. Chary for kindly providing us with output from his models.
P.G.~P.-G. also wishes to acknowledge support from the Spanish
Programa Nacional de Astronom\'{\i}a y Astrof\'{\i}sica under grant
AYA 2004-01676.

\bibliographystyle{apj}
\bibliography{referencias}

\appendix
\section{THE PHOTOMETRIC REDSHIFT TECHNIQUE}
\label{ppp}

Section~\ref{tech} presented the main characteristics of our
photometric redshift technique. Here we will describe the method in
detail and discuss the quality of the redshift estimations.

\subsection{Data compilation}

Our photometric redshifts benefited from the use of a vast amount of
data covering the UV, optical, NIR, and MIR spectral ranges. The main
characteristics of each dataset, including the wavelengths, limiting
magnitudes, and references for each filter, are given in
Tables~\ref{tableobservcdfs} and
\ref{tableobservhdfn}.

\placetable{tableobservcdfs}
\begin{deluxetable}{lrll|lrll|lrll}
\tabletypesize{\scriptsize}
\tablewidth{525pt}
\tablecaption{\label{tableobservcdfs}Characteristics of the data compiled for the CDFS.}
\tablehead{\colhead{Band}  & \colhead{$\lambda_{\mathrm{eff}}$} & \colhead{$m_{\mathrm{lim}}$}  & \colhead{Source} & \colhead{Band}  & \colhead{$\lambda_{\mathrm{eff}}$} & \colhead{$m_{\mathrm{lim}}$}  & \colhead{Source} & \colhead{Band}  & \colhead{$\lambda_{\mathrm{eff}}$} & \colhead{$m_{\mathrm{lim}}$}  & \colhead{Source}\\
\colhead{(1)} & \colhead{(2)} & \colhead{(3)} & \colhead{(4)} & \colhead{(1)} & \colhead{(2)} & \colhead{(3)} & \colhead{(4)} & \colhead{(1)} & \colhead{(2)} & \colhead{(3)} & \colhead{(4)}}
\startdata
MIPS-24       & 23.844  & 19.4 & {\it Spitzer} GTO & $v$           &  0.592  & 24.8 & GOODS$^c$   & $F485M$       &  0.486  & 25.2 & COMBO17$^e$ \\
IRAC-3.6      &  3.561  & 21.4 & {\it Spitzer} GTO & $i$           &  0.770  & 24.2 & GOODS$^c$   & $F518M$       &  0.519  & 25.0 & COMBO17$^e$ \\
IRAC-4.5      &  4.510  & 21.6 & {\it Spitzer} GTO & $z$           &  0.906  & 23.9 & GOODS$^c$   & $F571M$       &  0.572  & 24.9 & COMBO17$^e$ \\
IRAC-5.8      &  5.689  & 21.6 & {\it Spitzer} GTO & $J$           &  1.254  & 22.7 & GOODS$^c$   & $F604M$       &  0.605  & 24.7 & COMBO17$^e$ \\
IRAC-8.0      &  7.958  & 21.6 & {\it Spitzer} GTO & $H$           &  1.651  & 22.1 & GOODS$^c$   & $F646M$       &  0.645  & 24.4 & COMBO17$^e$ \\
$U$           &  0.365  & 24.2 & EIS$^a$           & $K$           &  2.161  & 22.0 & GOODS$^c$   & $F696M$       &  0.696  & 24.4 & COMBO17$^e$ \\
$U_p$         &  0.345  & 24.5 & EIS$^a$           & $J$           &  1.253  & 22.5 & EIS-DPS$^d$ & $F753M$       &  0.753  & 24.1 & COMBO17$^e$ \\
$B$           &  0.459  & 24.7 & EIS$^a$           & $K$           &  2.165  & 21.5 & EIS-DPS$^d$ & $F815M$       &  0.816  & 24.1 & COMBO17$^e$ \\
$V$           &  0.537  & 23.8 & EIS$^a$           & $U$           &  0.366  & 25.9 & COMBO17$^e$ & $F855M$       &  0.856  & 23.7 & COMBO17$^e$ \\
$R$           &  0.658  & 23.6 & EIS$^a$           & $B$           &  0.458  & 25.5 & COMBO17$^e$ & $F915M$       &  0.914  & 23.4 & COMBO17$^e$ \\
$I$           &  0.867  & 22.8 & EIS$^a$           & $V$           &  0.538  & 25.1 & COMBO17$^e$ & GALEX-fuv     &  0.152  & 24.5 & GALEX GTO   \\
$R$           &  0.658  & 24.4 & LCIS$^b$          & $R$           &  0.648  & 24.8 & COMBO17$^e$ & GALEX-nuv     &  0.231  & 24.4 & GALEX GTO   \\
$I$           &  0.810  & 24.0 & LCIS$^b$          & $I$           &  0.857  & 24.0 & COMBO17$^e$ & $I$, spectra  &  0.867  & 22.7 & VVDS$^f$    \\
$z$           &  0.901  & 23.4 & LCIS$^b$          & $F420M$       &  0.418  & 25.5 & COMBO17$^e$ &               &         &      &             \\
$b$           &  0.430  & 25.7 & GOODS$^c$         & $F464M$       &  0.462  & 25.2 & COMBO17$^e$ &               &         &      &             \\
\enddata
\tablecomments{(1) Name of the observing band. (2) Effective wavelength  
(in $\mu$m) of the filter calculated by convolving the Vega spectrum
\citep{1994AJ....108.1931C} with the transmission curve of the 
filter$+$detector. (3) Limiting AB magnitudes defined as the third
quartile of the magnitude distribution of our sample. (4) Source from
where the data were obtained: $^a$ ESO Imaging Survey (EIS,
\citealt{2002yCat..33790740A}); $^b$ Las Campanas Infrared Survey
(LCIS, \citealt{1999ASPC..191..148M}); $^c$ The Great Observatories
Origins Deep Survey (GOODS,
\citealt{2004ApJ...600L..93G}); $^d$ EIS Deep Public Survey
(EIS-DPS, \citealt{2001astro.ph..2300V}); $^e$ Classifying Objects by
Medium-Band Observations -a spectrophotometric 17-filter survey-
(COMBO17, \citealt{2004A&A...421..913W}); $^f$ VIRMOS-VLT Deep Survey
(VVDS, \citealt{2004A&A...428.1043L}).}
\end{deluxetable}

\placetable{tableobservhdfn}
\begin{deluxetable}{lrll}
\tabletypesize{\small}
\tablewidth{250pt}
\tablecaption{\label{tableobservhdfn}Characteristics of the data compiled for the HDFN.}
\tablehead{\colhead{Band}  & \colhead{$\lambda_{\mathrm{eff}}$} & \colhead{$m_{\mathrm{lim}}$}  & \colhead{Source}\\
\colhead{(1)} & \colhead{(2)} & \colhead{(3)} & \colhead{(4)}}
\startdata
MIPS-24       & 23.844  & 19.5 & {\it Spitzer} GTO \\
IRAC-3.6      &  3.561  & 21.6 & {\it Spitzer} GTO \\
IRAC-4.5      &  4.510  & 21.8 & {\it Spitzer} GTO \\
IRAC-5.8      &  5.689  & 21.8 & {\it Spitzer} GTO \\
IRAC-8.0      &  7.958  & 21.7 & {\it Spitzer} GTO \\
$U$           &  0.358  & 25.2 & Subaru Deep imaging$^a$\\
$B$           &  0.442  & 25.2 & Subaru Deep imaging$^a$\\
$V$           &  0.546  & 24.9 & Subaru Deep imaging$^a$\\
$R$           &  0.652  & 24.4 & Subaru Deep imaging$^a$\\
$I$           &  0.795  & 23.9 & Subaru Deep imaging$^a$\\
$z$   	      &  0.909  & 23.6 & Subaru Deep imaging$^a$\\
$b$           &  0.430  & 25.7 & GOODS$^b$ \\
$v$           &  0.592  & 24.9 & GOODS$^b$ \\
$i$           &  0.770  & 24.3 & GOODS$^b$ \\
$z$           &  0.906  & 23.9 & GOODS$^b$ \\
$HK_s$        &  2.127  & 21.3 & QUIRC Deep imaging$^a$\\ 
$b$, spectra  &  0.430  & 24.4 & TKRS$^c$ \\
\enddata
\tablecomments{(1)  Name of the observing band. (2) Effective wavelength  
(in $\mu$m) of the filter calculated by convolving the Vega spectrum
\citep{1994AJ....108.1931C} with the transmission curve of the 
filter$+$detector. (3) Limiting AB magnitudes defined as the third
quartile of the magnitude distribution of our sample. (4) Source from
where the data were obtained: $^a$ publicly available ultra-deep
optical and NIR data from
\citet{2004AJ....127..180C}; $^b$ The Great
Observatories Origins Deep Survey (GOODS,
\citealt{2004ApJ...600L..93G}); $^c$ Team Keck Treasury
Redshift Survey (TKRS, \citealt{2004AJ....127.3121W}) and
\citet{2004AJ....127.3137C}.}
\end{deluxetable}

\subsection{Models and fitting technique}

We built two sets of SEDs from the 24~\mic\, selected galaxies with
spectroscopic data in CDFS (redshifts from VVDS) and HDFN (redshifts
from TKRS and \citealt{2004AJ....127.3137C}). In both cases, we only
used the galaxies that were flagged as having accurate redshifts
(confidence higher than 80\%). We also selected only those galaxies
with more than ten data points in their SEDs, assuring that the
UV/optical, NIR, and MIR spectral ranges were covered.

The observed SEDs were deredshifted (taken to $z=0$) by transforming
the effective wavelengths of the filters (calculated with the filter
response and the detector quantum efficiency curves for each
observation dataset\footnote{From this point, every time we refer to
the filter response, we mean the filter transmission convolved with
the detector response.}) to the rest-frame and applying a $(1+z)$
factor to the flux density. The $z=0$ templates may suffer from
inadequate K-correction. Indeed, when we convolved the filter curves
with the $z=0$ templates and redshifted the results (to the original
redshift of the galaxy), we did not recover the original observed
SEDs. We used a minimization numeric method to obtain the best $z=0$
template compatible with the observed SED (i.e., we applied the
K-correction using this minimization algorithm).

%
%

\begin{figure}
\begin{center}
\includegraphics[width=11.0cm]{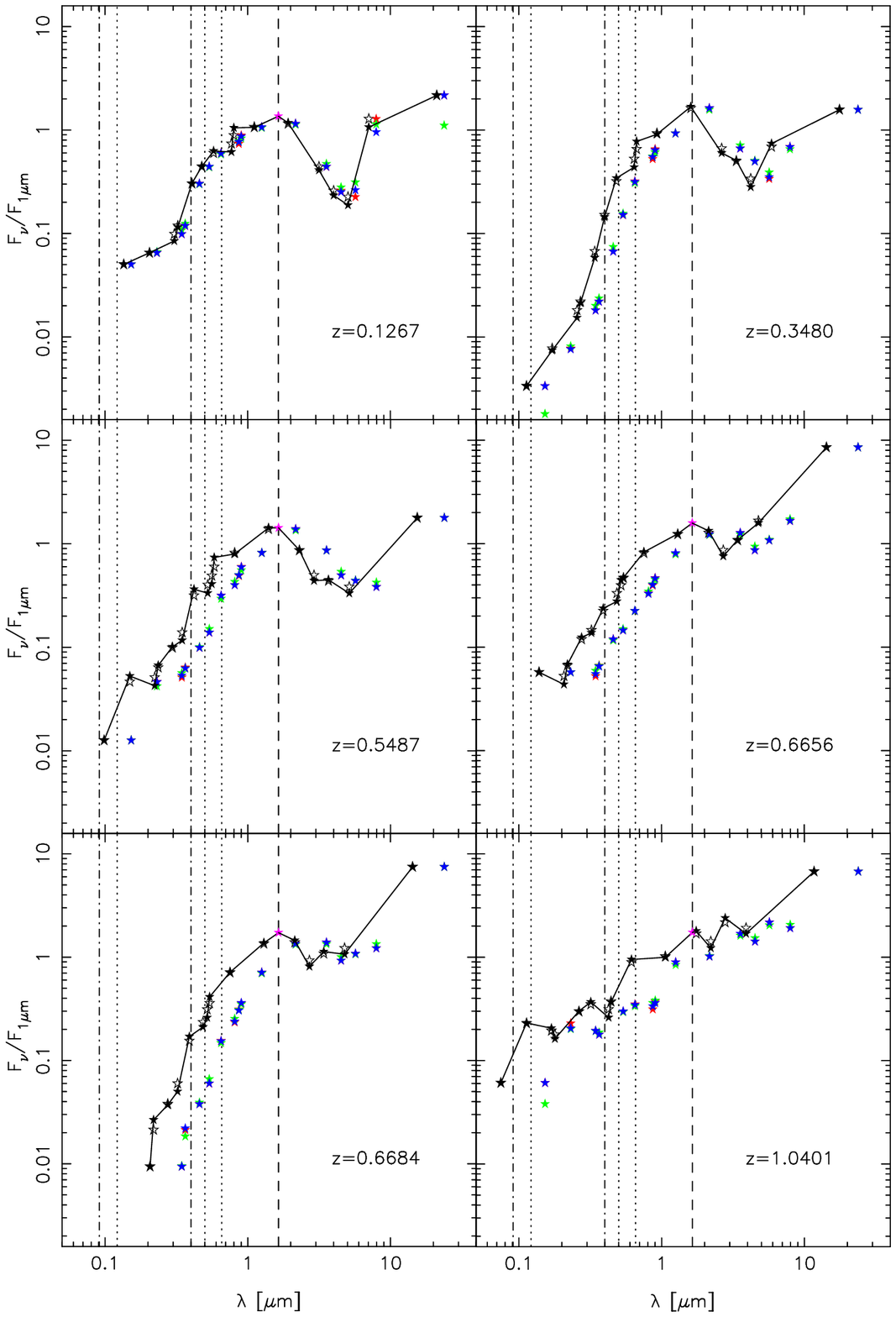}
\end{center}
\figcaption{\label{templas}Example templates obtained from the 
VVDS dataset (data normalized to the 1~\mic\, flux density). The
original redshift of the source is given for each galaxy. Red stars
show the original SED of the galaxy. Black open stars show the $z=0$
derived template without a K-correction calculation. Green stars show
the recovered SED built by convolution of the $z=0$ non-K-corrected
template with the filter response curves. Black filled stars (joined
by a black line) show the final $z=0$ template after applying the
K-correction. Blue stars show the final recovered redshifted template
(which must coincide with the template formed by red stars). In each
plot, the 1.6~\mic\, bump is marked with a dashed line (and a magenta
filled star), the Lyman and 4000~\AA\, breaks are marked with
dashed-dotted lines, and the positions of the Ly$\alpha$,
$[OIII]\lambda\lambda4959,5007$ and
H$\alpha+[NII]\lambda\lambda6548,6584$ emission-lines are marked with
dotted lines.}
\end{figure}

\slugcomment{Please, plot this figure in a single page}
\placefigure{templas}

We performed a visual inspection of all the templates, rejecting those
with unusual shapes due to deviant photometry points. The final
training sets were formed by 317 and 542 sources for the VVDS and HDFN
datasets, respectively. To these, we also added the
\citet{1999A&A...350..381D} 17 empirically-calibrated templates. 
Some representative examples of our templates are shown in
Figure~\ref{templas}. These examples show that the most prominent
feature of the great majority of the SEDs is the 1.6~\mic\, stellar
bump (marked with a dashed line). Nonetheless, for some galaxies this
feature is hardly visible (see the bottom-right template in
Figure~\ref{templas}). This absence of the stellar bump is observed in
some of the most infrared-luminous (presumably AGN-dominated) galaxies
in the local Universe \citep[see,
e.g.,][]{1988ApJ...328L..35S,1999A&A...350..381D}. Despite the low
resolution of the SEDs, some other spectral features are also visible,
such as emission-lines in the UV/optical (marked with dotted lines),
the 4000~\AA\, break (dashed-dotted line), or emission from PAHs in
the MIR. 

After building the set of empirical templates, we redshifted them to
values in the $0<z<3$ range using a step of $\Delta z=0.005$. The
redshift range was chosen based on expectations prior to launch
\citep{2003ApJ...585..617D}. These redshifted templates were then
convolved with the filter responses.

The redshift estimation proceeded as follows. First, the entire
observed SED was fitted with a Chebysev polynomial. When possible
(i.e., when there were NIR and IRAC data and the stellar bump was
present), the derivative of this polynomial was used to estimate the
position of the 1.6~\mic\, bump and its uncertainty. This position was
used as a first guess and to constrain the final solution (in the
range formed by the bump value and its error). We found this step to
be a good procedure to get rid of outliers. Second, the observed and
template fluxes were normalized to one of the bands (the reference
band mentioned in Section~\ref{merging}) to account for the effects of
different luminosities. Data points with dubious photometric
calibration or repeated observations (two observations for the same
filter) were removed before the fitting. In addition, the data points
at the edges of the SEDs (the bluest and reddest filters, the latter
always being 24~\mic), were given smaller weights in the fitting, or
removed when the templates had no data at those wavelengths. Third,
the templates and observed values were compared and a most probable
redshift was calculated by minimizing a reduced $\chi^2$ estimator of
the form:

\begin{equation}
\chi^2=\frac{1}{(N_\mathrm{filt}-1)}\sum_1^{N_\mathrm{filt}}\frac{(F^i_\mathrm{template}-F^i_\mathrm{observed})^2}{(\Delta F^i_\mathrm{observed})^2}
\end{equation}

\noindent where $N_\mathrm{filt}$ is the number of filters considered, 
$F_\mathrm{template}$ is the flux calculated for each redshifted
template in the $ith$ filter, and $F_\mathrm{observed}$ and $\Delta
F^i_\mathrm{observed}$ are the measured fluxes and uncertainties in
each filter. Errors in the redshift were calculated with a
$\Delta\chi^2$ algorithm, and were quoted as the $z$-range for which
the solution has a 68\% probability of being correct. We only obtained
redshifts for galaxies with more than four points in the SED
(virtually all the sources mentioned in Section~\ref{merging}).

\subsection{Comparison with the spectroscopic sample}

\begin{figure}
\begin{center}
\includegraphics[angle=-90,width=8cm]{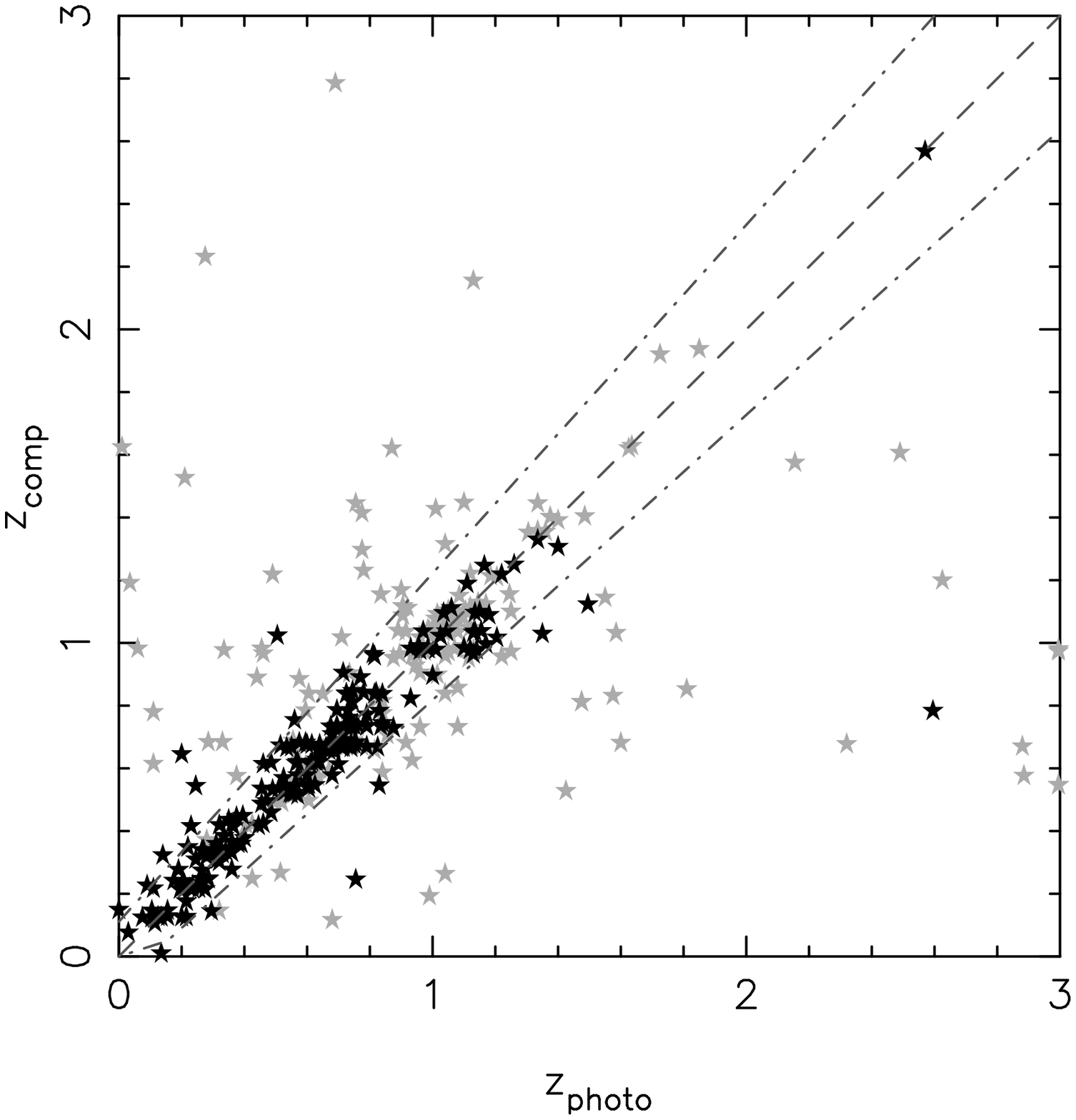}
\includegraphics[angle=-90,width=8cm]{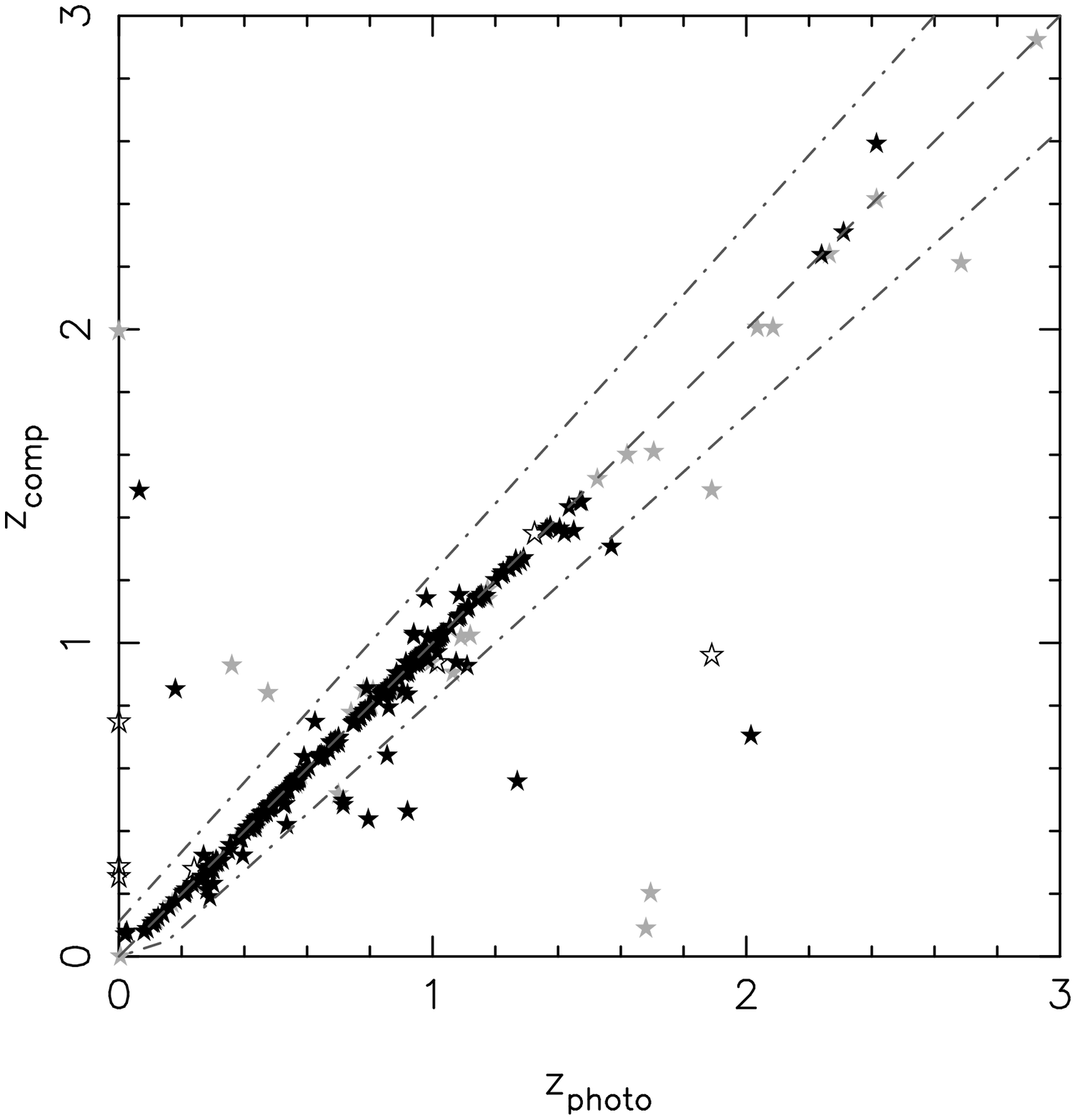}
\figcaption{\label{comp_spec}Comparison between the photometric 
redshifts obtained in this work and the spectroscopic redshifts
measured in CDFS (left panel) and in HDFN (right panel). Both
comparisons refer to the photometric redshift results obtained with
the complete template set (built with 317 sources from VVDS, 542 from
TKRS and \citealt{2004AJ....127.3137C}, and the 17 templates in
\citealt{1999A&A...350..381D}). Gray symbols are sources with 
unreliable spectroscopic redshifts. Open stars are sources detected in
less than five bands. The dashed line shows the equality line, and the
dash-dotted ones show the $\sigma_z/(1+z)<0.1$ area.}
\end{center}
\end{figure}

\slugcomment{Please, plot this figure with the width of two columns}
\placefigure{comp_spec}

The redshifts obtained with our photometric technique were compared
with the spectroscopic values for all the galaxies in the sample. In
Figure~\ref{comp_spec}, we demonstrate that our procedure is able to
recover the redshifts of the galaxies used as an input for the
technique. 

The left panel in this Figure shows the comparison of photometric and
spectroscopic redshifts for CDFS when using the complete template set
(876 templates in total). The distribution of points is very symmetric
around the equality line. The median value for the difference between
the derived photometric redshift and the spectroscopic one ($\delta z$
hereafter) is $\delta z=0.003$, showing there is no systematic
difference. Given that the wavelength scales in terms of $(1+z)$ when
going to more distant sources, it seems more physically meaningful to
discuss errors in terms of $\sigma_z/(1+z)$
\citep{2004A&A...421..913W}, where $\sigma_z$ is the absolute value of
$\delta z$. For the VVDS comparison, 90\% of the objects have values
of $\sigma_z/(1+z)<0.2$, 82\% of the objects have values of
$\sigma_z/(1+z)<0.1$, and 75\% have $\sigma_z/(1+z)<0.05$. The average
(median) $\sigma_z/(1+z)$ is 0.050 (0.012). These statistics improve
if we only take into account sources with highly reliable redshifts
(as stated by VVDS): 98\% of the objects have values of
$\sigma_z/(1+z)<0.2$, 92\% of the objects have values of
$\sigma_z/(1+z)<0.1$, and 85\% have $\sigma_z/(1+z)<0.05$.

The right panel of Figure~\ref{comp_spec}\, shows the comparison of
our photometric redshifts with spectroscopic redshifts in the HDFN
when using the complete template set. The statistics for this
comparison (for the highly-reliable spectroscopic sample) are very
similar to the CDFS case: $<\delta z>=-0.003$,
$<\sigma_z/(1+z)>=0.016$, 98\% of the objects have values of
$\sigma_z/(1+z)<0.2$, 97\% of the objects have values of
$\sigma_z/(1+z)<0.1$, and 95\% have $\sigma_z/(1+z)<0.05$. It is
interesting to notice the smaller scatter of the points around the
equality line in comparison with what we found in CDFS. This
difference is due to the larger number of sources having NIR data in
the HDFN, which makes the photometric redshift estimation more
reliable because the 1.6~\mic\, bump position can be better
constrained.

The two panels in Figure~\ref{comp_spec} show some outliers, a total
of 75 objects with $\sigma_z/(1+z)>0.1$ for the left panel (18\% of
the total sample of 425 sources in CDFS), and 22 objects with
$\sigma_z/(1+z)>0.1$ for the right panel (4\% of the total sample of
601 sources in HDFN). We will discuss these photometric redshift
failures next, in order to characterize why the photometric redshift
technique fails. None of these outliers were included within the
template set, either because of a unreliable spectroscopic redshift,
or because they did not have more than ten points in their SEDs.

In CDFS, 60 out of the 75 outliers (i.e., 80\% of the outliers) do not
have highly-reliable spectroscopic redshifts (gray stars in
Figure~\ref{comp_spec}). Out of these, 26 present slight differences
with the spectroscopic redshifts of $\sigma_z/(1+z)<0.15$ (normally
the redshift is overestimated). Out of the 34 remaining sources with
unreliable spectroscopy, 18 present a clearly wrong photometric
redshift. Half of these 18 sources present power-law like SEDs, and
our method fails to obtain a good redshift. The other half do not
present any particular problem, but the template selected by our
technique gives a wrong redshift. The 16 remaining sources with
non-reliable spectroscopy and $\sigma_z/(1+z)>0.15$ show SEDs clearly
incompatible with the quoted spectroscopic redshifts. Out of the total
75 outliers in the left panel of Figure~\ref{comp_spec}, 15 are
secured spectroscopic identifications, with 10 of them presenting
$\sigma_z/(1+z)<0.15$. The other 5 do not have IRAC photometry, and
our technique gives a very large redshift for them.

In HDFN, 4 out of the 22 outliers have less than 5 points in their
SEDs. This kind of object was removed from the photometric redshift
sample in this paper due to the high uncertainties related to the
small number of data points. In fact, there are a total of 7 galaxies
within this group with reliable spectroscopic redshift in the entire
HDFN sample, which gives more than 50\% (4 out of 7) having a wrong
photometric redshift. There are 8 more outliers with unreliable
spectroscopic redshifts. Out of these, 3 present slightly
overestimated photometric redshifts, with differences with the
spectroscopic redshifts of $\sigma_z/(1+z)<0.15$. Another 2 sources
(out of the 8 with unreliable spectroscopy) present a clearly wrong
photometric redshift (one of the gray stars at $z\sim1.7$, and the
gray star at $z_\mathrm{photo}=0$) due to highly deviant points in the
SEDs (e.g., the $V$-band flux is 10 times larger than what would be
expected based on other adjacent data points, the $BvR$ bands, in the
$z_\mathrm{photo}\sim1.7$ case). This deviant point could be linked to
source variability or deblending problems. The 3 remaining sources
with unreliable spectroscopy show SEDs clearly incompatible with the
quoted spectroscopic redshifts. Out of the other 10 outliers found in
the right panel of Figure~\ref{comp_spec} (and having reliable
spectroscopy), 4 of them present $\sigma_z/(1+z)<0.15$ (with a
slightly overestimated photometric redshift). Another 6 are clearly
wrong photometric redshifts, 3 of them lacking for NIR and/or IRAC
data, and the other 3 objects presenting very disturbed SEDs.

In summary, the outliers in the two panels of Figure~\ref{comp_spec}
are sources with wrong photometric redshifts due to photometry
problems (about one third of them), objects with slightly
overestimated redshifts ($\sim$50\% of them), or probably wrong
spectroscopic redshifts (10\%--20\%). This demonstrates that the
fitting method does not introduce significant errors in the redshift
determination.

\begin{figure}
\begin{center}
\includegraphics[angle=-90,width=8cm]{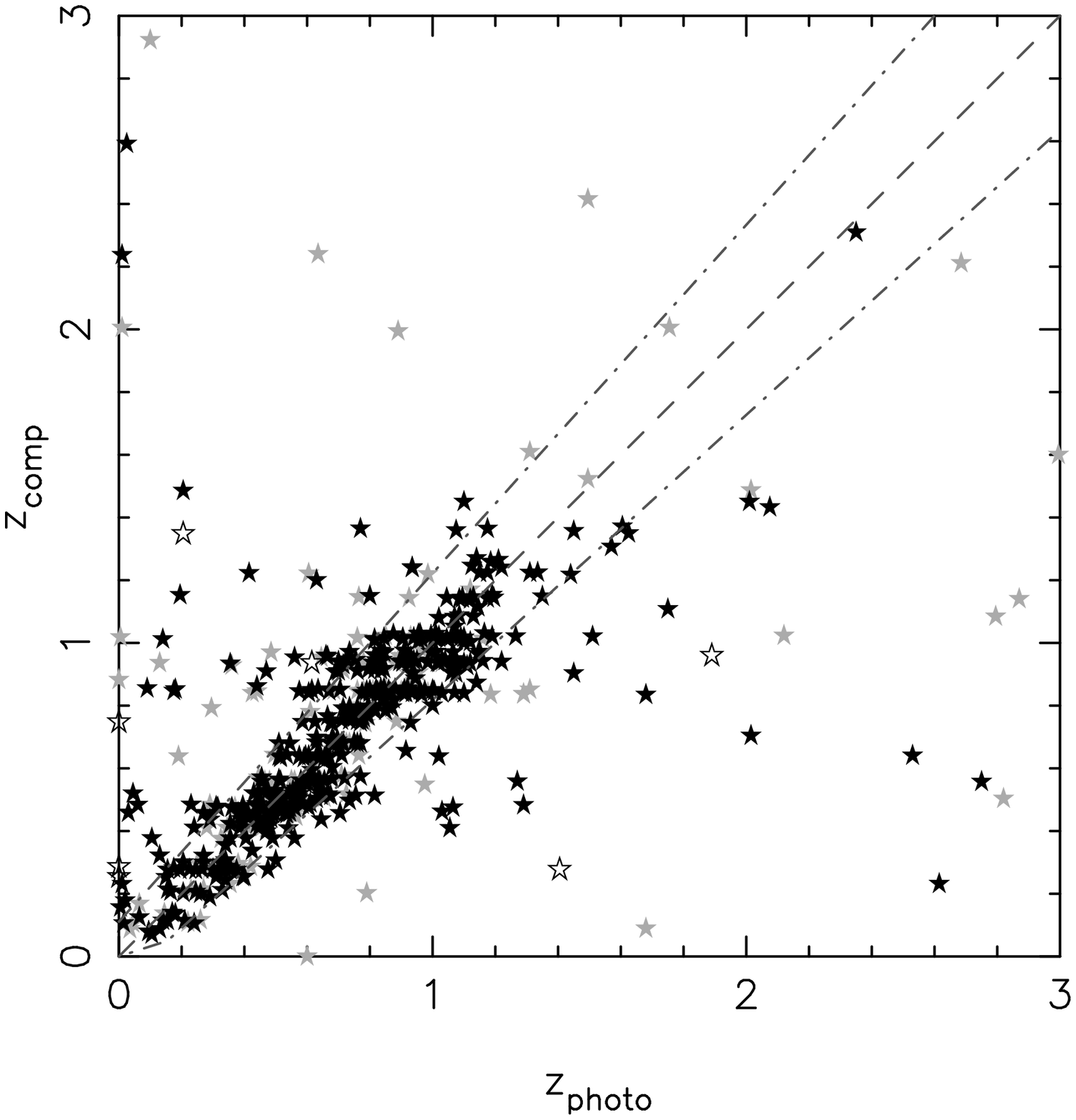}
\includegraphics[angle=-90,width=8cm]{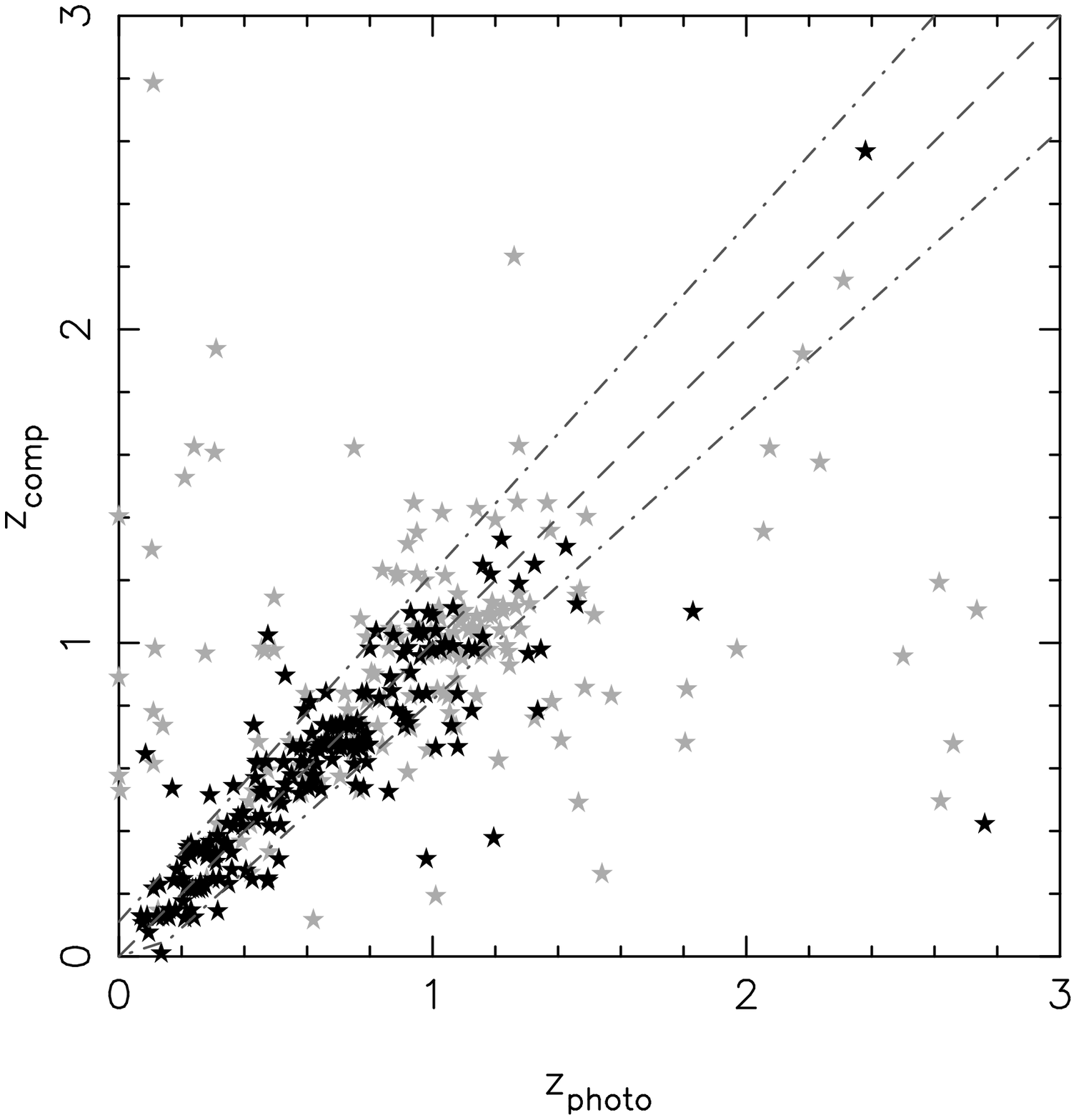}
\figcaption{\label{phz}{\it Left:} Comparison between the photometric 
redshifts obtained in this work and the spectroscopic redshifts for
HDFN sources. This comparison refers to the photometric redshift
results obtained by using the templates built in CDFS (317
templates). {\it Right:} Comparison between the photometric redshifts
obtained in this work and the spectroscopic redshifts for CDFS
sources. This comparison refers to the photometric redshift results
obtained by using the templates built in HDFN (542 templates). For
both panels, gray symbols are sources with unreliable spectroscopic
redshifts. Open stars are sources detected in less than five
bands. The dashed line shows the equality line, and the
dash-dotted ones show the $\sigma_z/(1+z)<0.1$ area.}
\end{center}
\end{figure}

\slugcomment{Please, plot this figure with the width of two columns}
\placefigure{phz}

Most of the galaxies plotted in the left panel (more precisely, 60\%)
and right panel (90\%) of Figure~\ref{comp_spec} were included in the
template set used to derive photometric redshift. Therefore, the
numbers given are not entirely representative of the goodness of the
method presented in this paper. A real test of the method is presented
in Figure~\ref{phz}. Here we divided the sample in roughly two
parts. One half of the sample was formed by the sources in CDFS, and
the other half by sources in HDFN. We then obtained photometric
redshifts for all the sources in one half of the sample by using the
templates built from sources with a spectroscopic redshift within the
other half of the sample.  The left panel of Figure~\ref{phz} shows
the comparison of photometric and spectroscopic redshifts for the HDFN
sources when using the templates built from CDFS sources. The right
panel in Figure~\ref{phz} shows the comparison for CDFS sources when
only using templates from HDFN.  The Figure demonstrates that we are
able to obtain photometric redshifts (for the sources with secure
spectroscopic redshift) with $\sigma_z/(1+z)<0.1$ for at least 80\% of
the sample (80\% for the left panel and 84\% for the right one), and
redshifts with $\sigma_z/(1+z)<0.2$ for more than 90\% of the galaxies
(91\% for the left panel and 94\% for the right one). Other statistics
are: $<\delta z>=-0.001$, $<\sigma_z/(1+z)>=0.078$ for the left panel
and $<\delta z>=-0.002$, $<\sigma_z/(1+z)>=0.077$ for the right panel.

We also analyzed all the outliers on Figure~\ref{phz} to characterize
why the photometric redshift technique fails. In the left panel, there
are 601 galaxies with spectroscopic redshifts, 125 (21\% of the total
spectroscopic sample in HDFN) of which present
$\sigma_z/(1+z)>0.1$. Out of this number, 67 sources (54\% of the
outliers) have $\sigma_z/(1+z)<0.2$, half of them with overestimated
photometric redshifts, half with underestimated values. Within the 58
sources remaining (from the 125 outliers), 7 galaxies present very
disturbed SEDs (6\% of the outliers). Another 36 objects are clear
photometric redshift errors (29\% of the outliers), probably because
of the lack of NIR and/or IRAC data (13 objects out of the 36), or
because they present power-law SEDs (10 objects). The rest, 15 sources
(12\% of the outliers), are probable spectroscopic redshift errors
(virtually all of them flagged as unreliable spectroscopic
estimations). In the right panel of Figure~\ref{phz}, there are 425
galaxies, 81 (19\% of the total spectroscopic sample in CDFS) which
present $\sigma_z/(1+z)>0.1$. Out of this number, 25 (31\% of the
outliers) have $\sigma_z/(1+z)>0.2$, and only 10 of those are labeled
as secure spectroscopic redshifts, all with very disturbed SEDs. In
summary, less than 20\% of the galaxies present $\sigma_z/(1+z)>0.1$,
at least one third of those present $\sigma_z/(1+z)>0.2$, and from
these 10\%, at least one fourth are probable spectroscopic redshift
errors. Most of the sources with $\sigma_z/(1+z)>0.1$ either lack for
NIR/IRAC photometry, or present disturbed SEDs probably linked to
source variability (related to AGN activity) or deblending
problems. In fact, $\sim$6\% of the sources with $\sigma_z/(1+z)<0.1$
present multiple identifications (within the search radius), in
comparison with a $\sim$10\% for galaxies with $\sigma_z/(1+z)>0.1$.

\begin{figure}
\begin{center}
\includegraphics[angle=-90,width=9cm]{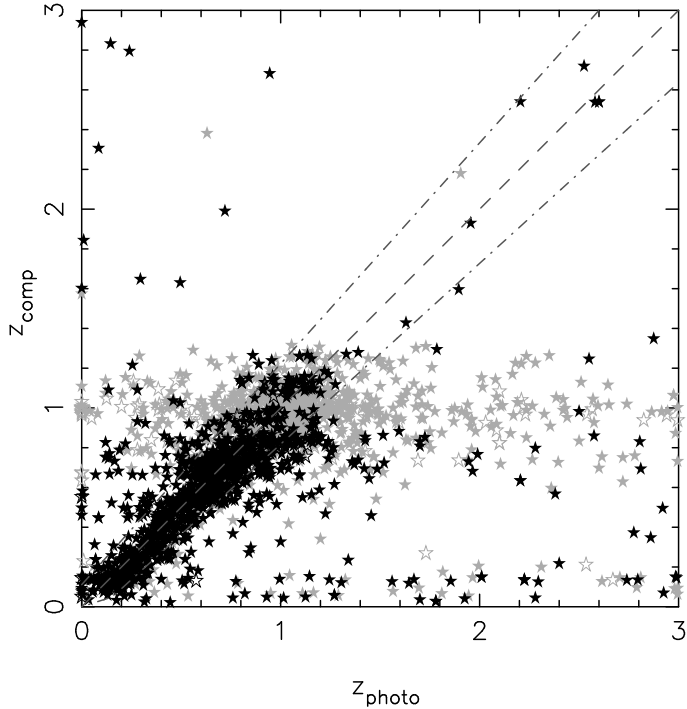}
\figcaption{\label{combo17} Comparison between the photometric 
redshifts obtained in this work (using the complete template set) and
the ones obtained by COMBO17 \citep{2004A&A...421..913W} in CDFS. Gray
symbols are sources with $\Delta z>0.05$ as given by COMBO17. Open
stars are sources detected in less than five bands. The dashed line
shows the equality line, and the dash-dotted ones show the
$\sigma_z/(1+z)<0.1$ area.}
\end{center}
\end{figure}

\slugcomment{Please, plot this figure with the width of one column}
\placefigure{combo17}

Finally, we also compared our results with other photometric redshift
surveys such as COMBO17 \citep{2004A&A...421..913W}. COMBO17 and our
redshifts agree very well up to $z\sim1$, with more than 90\% of the
sources being within the $\sigma_z/(1+z)<0.1$ area (just for a {\it
highly-reliable} sample, i.e., sources with $\Delta z<0.05$ as given
by COMBO17). However, there are two issues in the comparison, that we
can clearly see in Figure~\ref{combo17}. First, the distribution of
points is not symmetric at $z\gtrsim1$. Many sources lying near the
COMBO17 redshift limit ($z\sim1.4$) are placed at higher redshifts by
our photometric redshift method, most of them presenting high COMBO17
redshift uncertainties. For $0<z_{COMBO17}<1.4$, we find: $<\delta
z>=0.052$, $<\sigma_z/(1+z)>=0.114$, and 75\% of the sources have
$\sigma_z/(1+z)<0.1$. COMBO17 is known to have a deficiency when
estimating redshifts at $z\geq1$, because the useful spectral features
go out of their optical filter set. These are also the faintest
sources in the COMBO17 sample, and the photometric redshift method
gets more uncertain. In our case, NIR data are available for 50\% of
the sources, and IRAC photometry for virtually all of them. These
bands allow us to trace the spectral features used by COMBO17 to high
redshift ($z\sim3$), and to break redshift degeneracies coming from
the misidentification of the Balmer break with the Lyman break. The
second issue seen in Figure~\ref{combo17} is that some galaxies
(approximately 5\% of the common sources between our survey and
COMBO17) are placed at $z<0.2$ by COMBO17 and at higher redshifts by
our work. \citet{emeric05} also noticed this effect in studying the
luminosities of MIPS sources in CDFS, concluding that they must be
redshift outliers in the COMBO17 survey.

The statistics and the fraction of outliers of our photometric
redshift technique are comparable to most other photometric redshift
works in the literature (e.g., \citealt{1995AJ....110.2655C,
2003ApJ...586..745C,2003MNRAS.339.1195F,2003MNRAS.345..819R,
2004MNRAS.353..654B,2004ApJS..155...73Z}), and only slightly worse
than some surveys aimed at obtaining high-quality photometric redshift
by using adequate sets of narrow-band filters, although the new {\it
Spitzer} data helps to obtain more reliable and accurate redshifts for
$z\gtrsim1$ sources (e.g., \citealt{2004A&A...421..913W}).


Although \citet{2004ApJS..154..170L} demonstrated the ability of
\spitzer\, to identify sources in the 'redshift 
desert' \citep[$1.5<z<3.0$, ][] {2004ApJ...604..534S}, our photo-$z$
technique might have deficiencies for this redshift range. The
spectroscopic surveys are highly biased towards $z<1$ galaxies, given
their technical limitations for very faint objects. Indeed, our
template set does not have many galaxies lying at high redshifts: 80\%
of the templates correspond to $z<1$ sources, 16\% to $1.0<z<1.4$
galaxies, and only 4\% above $z=1.4$. Our technique assumes that the
most distant galaxies in our sample are very similar (in their SED
properties) to the closer ones, i.e., we can find templates fitting
the $z>1.4$ galaxies among the $z<1.4$ sample. Figure~\ref{zlum} shows
that this extrapolation is not unreasonable, given that the LIRG
population (probably showing similar properties at all redshifts)
dominates our sample (in number of sources) from $z\sim0.8$ up to
$z\sim2.0$. Only above $z\sim2.0$, the ULIRGs are statistically
relevant in number, and we do not have many templates with known
spectroscopic redshifts. 

\begin{figure}
\begin{center}
\includegraphics[angle=-90,width=12cm]{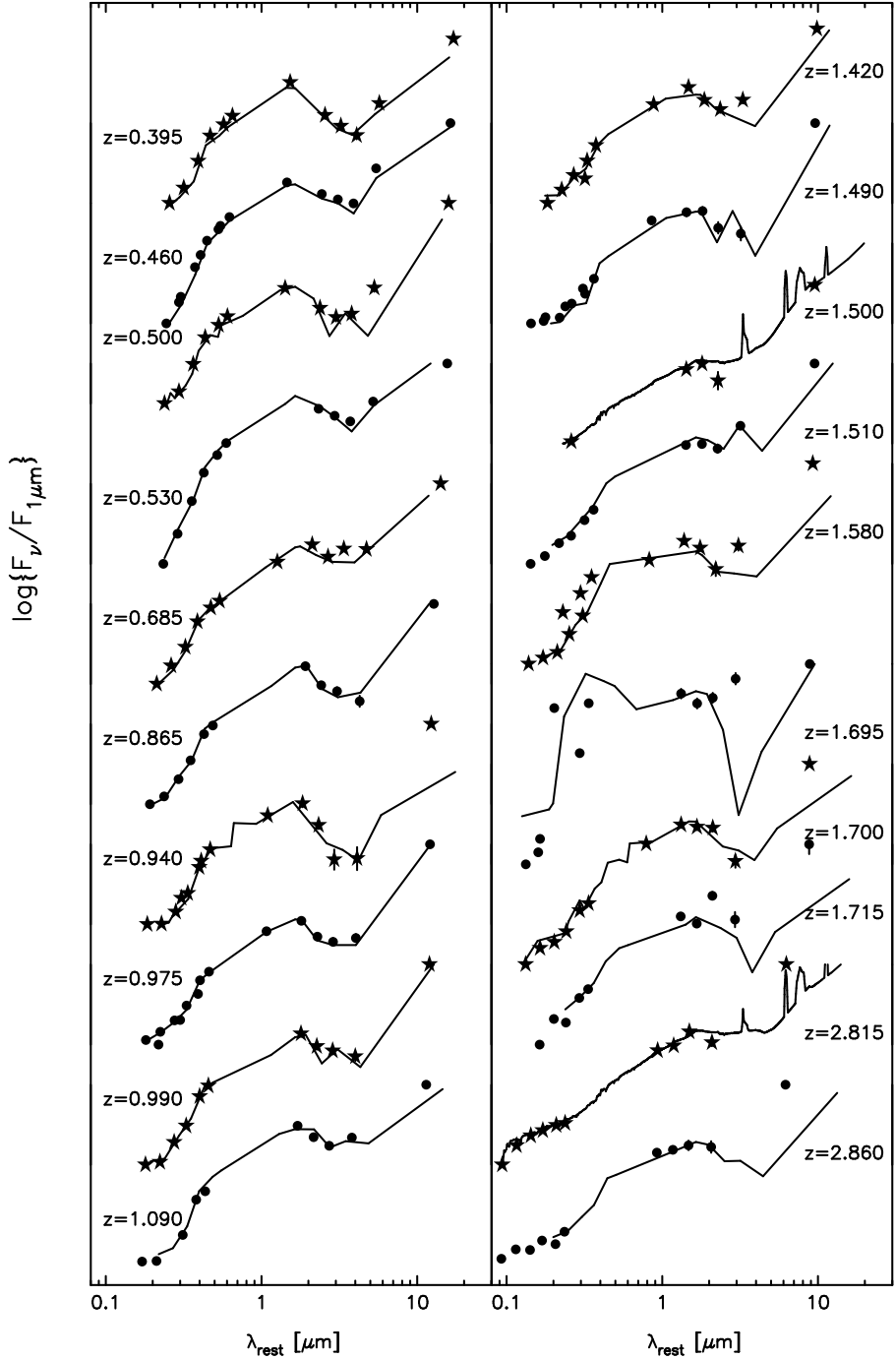}
\figcaption{\label{fits}Some randomly-selected examples of the fits 
obtained in the photometric redshift estimation. On the left, we show
ten sources selected randomly in the redshift range $0<z<1.4$, and on
the right ten more randomly selected sources at $1.4<z<3.0$ are
given. Photometry points are plotted with stars, and the best template
fit to the data is shown with a solid line.}
\end{center}
\end{figure}

Having this possible problem in mind, we performed a visual inspection
of the fits for the galaxies identified as being at $z>1.4$ by our
photo-$z$ technique. This test is represented in Figure~\ref{fits},
which plots 10 sources selected randomly from the entire sample in the
redshift range $0<z<1.4$ (left column), and 10 more at $1.4<z<3.0$
(right column). This figure gives an overview of the reliability of
our photo-$z$ technique. Most of the galaxies in our sample show a
clear 1.6~\mic\, stellar bump, a feature that allows a reliable
photometric redshift determination. Indeed, all $z<1.4$ galaxies in
Figure~\ref{fits} show a negative average slope in the IRAC bands, and
a positive slope in the optical/NIR bands, pointing to a clear
1.6~\mic\, bump. The determination of the exact position of this bump
is, however, importantly affected by the availability of NIR data (and
consequently, the uncertainty in the photometric redshift). For the
sources at $z>1.4$, at least 7 galaxies present a change in slope
(from positive to negative) inside the spectral range covered by IRAC,
which indicates the presence of the 1.6~\mic\, bump (for a
$1.3\lesssim z\lesssim 4$ galaxy). The visual inspection of the
randomly selected sources in Figure~\ref{fits} revealed that the
assigned redshift was dubious for 20\%--25\% of the sources, all of
them lacking a marked 1.6~\mic\, stellar bump (probably linked to the
presence of an AGN), and preferentially lying at $z>1.4$. This
percentage is very similar to the total reliability of our technique
($\sim80$\%), previously discussed in this Section, which suggests
that the procedure remains applicable to the highest redshift range. A
better coverage of the 'redshift desert' for LIRGs and ULIRGs by
spectroscopic surveys would be desirable to reduce the uncertainty of
the results achieved at $z>1.4$.

\slugcomment{Please, plot this figure in a single page}
\placefigure{fits}

\section{SYSTEMATIC UNCERTAINTIES IN THE LUMINOSITY FUNCTION
ESTIMATION LINKED TO PHOTOMETRIC REDSHIFT ERRORS}
\label{montecarlo}

In this Appendix, we will investigate the effect of the photometric
redshift errors discussed in Appendix~\ref{ppp} (i.e., scatter and
fraction of outliers) in the estimation of the luminosity functions.

The photometric redshift errors propagate in the calculation of
luminosities. Given that the estimation of the luminosity functions
involves binning and averaging of luminosities, the uncertainties
linked to redshift errors could in principle diminish, and they should
most probably be random. However, there could also exist some
systematic uncertainties, specially in the extremes of the luminosity
function (faint and bright ends), where the number of detected
galaxies is small, and the redshift outliers are preferentially found.

To understand the uncertainties in the luminosity function linked to
photometric redshift errors, we performed a Monte Carlo simulation
similar to the one presented in
\citet{2003ApJ...586..745C}. This simulation consisted in the
generation of an artificial catalog of galaxies following an assumed
input Schechter luminosity function and presenting the characteristic
limiting fluxes of our survey. Each galaxy was assigned a random
redshift between $z=0$ and $z=3$. This redshift was perturbed by an
amount linked to the probability distribution built from the scatter
of points in the comparison between spectroscopic and photometric
redshifts presented in Figure~\ref{phz}. This probability distribution
of photometric redshift uncertainties accounts for the scatter of
points around the equality line in Figure~\ref{phz}, and for the
outliers (i.e., the most deviant points in that scattering plot,
caused by photometry problems or any other issue). Finally, we
estimated the luminosity function for three representative redshift
intervals: $0.0<z<0.2$, $0.8<z<1.0$, and $1.8<z<2.2$. We will
concentrate our discussion in the lowest redshift interval,
$0.0<z<0.2$, where we can explore the largest luminosity range. Note
that this interval only probes the photometric redshift errors toward
higher redshifts. Given that we obtained very similar results for the
other intervals, we concluded that the errors toward lower redshifts do
not affect the results from our simulation significantly.

\begin{figure}
\begin{center}
\includegraphics[width=11.0cm,angle=-90]{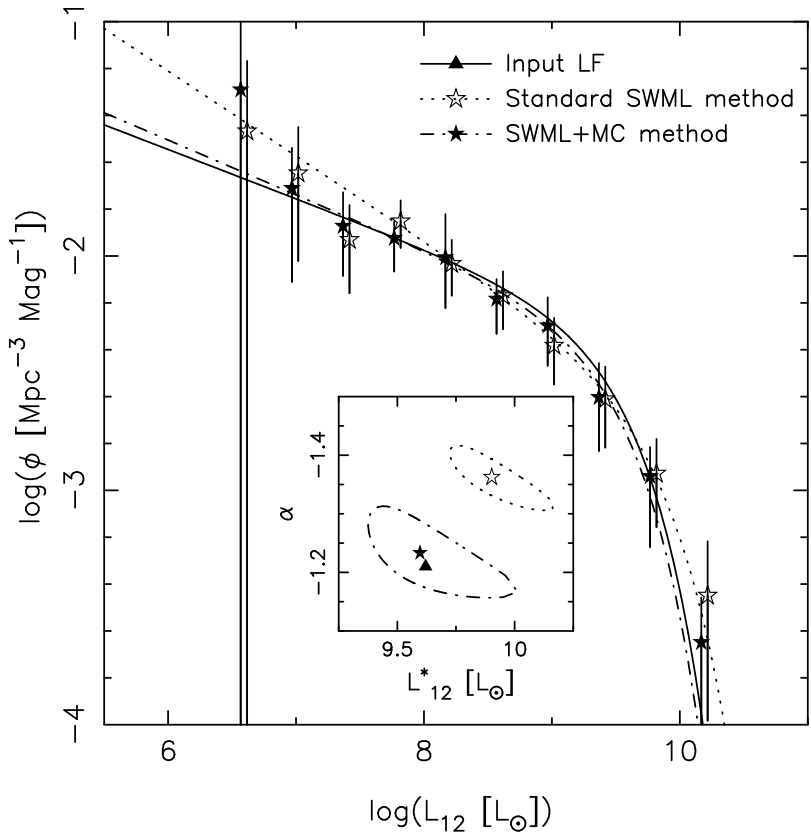}
\end{center}
\figcaption{\label{mc_sim}Results from the Monte Carlo simulation 
performed to investigate the systematic uncertainties in the 12~\mic\,
luminosity function estimation introduced by the photometric redshift
errors. The solid line shows the input Schechter luminosity function,
whose $\alpha$ and $L^*$ values are indicated by the triangle in the
inset. The open stars and dotted line (also in the inset, with the
contour delimiting the 99.99\% probability area) show the recovered
luminosity function for the artificial catalog. These open stars have
been offset to higher luminosities by a constant small amount for
clarity. The filled stars and dash-dotted line (also in the inset)
represent the final luminosity function obtained with the modified
SWML method (using a Monte Carlo iterative method). This technique
accounts for the typical luminosity function errors (linked to the
number of galaxies in each luminosity bin), as well as for redshift
uncertainties.}
\end{figure}

The results from the Monte Carlo simulation are plotted in
Figure~\ref{mc_sim}. The input luminosity function (continuous line)
is not well recovered with the standard SWML method (open stars fitted
by the dotted line). As we previously suggested, the photometric
redshift errors affect the faint and bright ends of the luminosity
function, resulting on an overestimation of the galaxy density in
these ranges, which turns into an overestimation of both $\alpha$ (by
$\sim$20\%) and $L^*$ (by $\sim$0.3~dex). This result is consistent
with that found by \citet{2003ApJ...586..745C}. Note, however, that
although the luminosity function parameters are not properly recovered
due to photometric redshift errors, the integrated luminosity density
does not change significantly (less than 0.1~dex, below the estimated
uncertainty 0.25~dex).

\slugcomment{Please, plot this figure with the width of one column}
\placefigure{mc_sim}

To cope with the previously described systematic errors, we modified
the SWML luminosity function method to include the photometric
redshift and luminosity uncertainties. We used a Monte Carlo approach
where each redshift was treated as statistical variable with an
average and an uncertainty. The average was the photometric redshift
given by our technique and the uncertainty was derived from the
probability distribution of redshift errors extracted from
Figure~\ref{phz}. The redshifts for the whole catalog of galaxies in
our survey were randomly perturbed according to this distribution, and
the 12~\mic\, luminosities and their uncertainties were
recalculated. The luminosity function was estimated using the SWML
method for the new catalog. We iterated this process 1000 times, and
the final luminosity function points (given in Figures~\ref{locallf}
and \ref{lf0t}), the fitting parameters, and the associated errors for
all these quantities were derived from the distribution of solutions.

The results from the modified SWML method for the simulation described
previously are shown in Figure~\ref{mc_sim} with filled stars and a
dash-dotted line (also in the inset). We are able to recover the input
luminosity function parameters with high accuracy, reducing the
systematic errors to less than 2\% for $\alpha$, and less than
0.04~dex for both $L^*$ and $\phi^*$. In general, the modified SWML
method obtained luminosity functions with smaller faint-end slopes and
lower $L^*$ values than the traditional SWML technique for all the
redshift ranges.

\end{document}